\documentclass[groupedaddress,a4paper,showpacs,showkeys,nofootinbib,preprint,tightenlines]{revtex4}
\usepackage{amsmath,amsthm,amssymb}
\usepackage[paper=a4paper,margin=3cm]{geometry}
\newcommand{\al}{\alpha}
\newcommand{\be}{\beta}
\newcommand{\de}{\delta}
\newcommand{\ep}{\epsilon}
\newcommand{\vep}{\varepsilon}
\newcommand{\ga}{\gamma}

\newcommand{\la}{\lambda}
\newcommand{\om}{\omega}
\newcommand{\si}{\sigma}

\newcommand{\vp}{\varphi}
\newcommand{\ze}{\zeta}
\newcommand{\La}{\Lambda}
\newcommand{\Si}{\Sigma}
\newcommand{\bx}{\mathbf{x}}
\newcommand{\bsi}{\boldsymbol{\si}}
\newcommand{\btau}{\boldsymbol{\tau}}

\newcommand{\bs}{\mathbf{s}}

\newcommand{\bz}{\mathbf{z}}

\newcommand{\Th}{\tilde{h}}

\newcommand{\tcH}{\widetilde{\cH}}

\newcommand{\ts}{\tilde{s}}
\newcommand{\tPhi}{\widetilde{\Phi}}
\newcommand{\ssH}{\mathsf{H}}

\newcommand{\hq}{\hat{q}}
\newcommand{\hs}{\hat{s}}

\newcommand{\hcH}{\widehat{\mathcal H}}

\newcommand{\hPhi}{\widehat{\Phi}}
\newcommand{\hPsi}{\widehat{\Psi}}
\def\CC{\mathbb{C}}
\def\NN{\mathbb{N}}

\def\ZZ{\mathbb{Z}}

\newcommand{\cB}{{\mathcal B}}

\newcommand{\cH}{{\mathcal H}}

\newcommand{\cJ}{{\mathcal J}}
\newcommand{\cL}{{\mathcal L}}

\newcommand{\cP}{{\mathcal P}}

\newcommand{\cS}{{\mathcal S}}
\newcommand{\cT}{{\mathcal T}}
\newcommand{\cX}{{\mathcal X}}

\newcommand{\fsl}{\mathfrak{sl}}

\def\Bx{\bar x}

\def\BH{\,\overline{\!H}{}}
\newcommand{\BcH}{\overline{\cH}{}}

\def\BP{\,\overline{\!P}{}}
\def\BQ{\,\overline{\!Q}{}}
\def\BR{\,\overline{\!R}{}}

\newcommand{\pa}{\partial}

\def\ket#1{|#1\rangle}

\def\BStrut{\vrule height12pt depth6pt width0pt}
\let\ds\displaystyle
\let\ni\noindent
\newcommand{\ms}{\mspace{1mu}}
\newcommand{\iu}{{\mathrm i}}
\newcommand{\e}{{\mathrm e}}
\newtheorem{theorem}{\bf Theorem}
\newtheorem{lemma}{\bf Lemma}
\newtheorem{corollary}{\bf Corollary}
\newtheorem{proposition}{\bf Proposition}
\newcounter{rem}
\newenvironment{remark}{\medskip{\ni\em Remark~\therem.}}{\newline\stepcounter{rem}}
\newcounter{ex}
\newenvironment{example}{{\ni\em Example~\theex.}}{\newline\stepcounter{ex}}

\begin{document}
\title{Exchange operator formalism for $N$-body spin models\\
with near-neighbors interactions}
\author{A. Enciso}
\author{F. Finkel}
\author{A. Gonz{\'a}lez-L\'opez}
\author{M.A. Rodr{\'\i}guez}

\affiliation{Departamento de F{\'\i}sica Te\'orica II, Universidad
Complutense de Madrid, 28040 Madrid, Spain}

\date{January 10, 2007}

\begin{abstract}
We present a detailed analysis of the spin models with
near-neighbors interactions constructed in our previous
paper~\cite{EFGR05b} by a suitable generalization of the exchange
operator formalism. We provide a complete description of a certain
flag of finite-dimensional spaces of spin functions preserved by
the Hamiltonian of each model. By explicitly diagonalizing the
Hamiltonian in the latter spaces, we compute several infinite
families of eigenfunctions of the above models in closed form in
terms of generalized Laguerre and Jacobi polynomials.
\end{abstract}

\maketitle

\section{Introduction}\label{sec.intro}

The discovery of the quantum models
named after Calogero~\cite{Ca71} and Sutherland~\cite{Su71,Su72}
is a key development in the theory of integrable systems
which has exerted a far-reaching influence on many different areas
of Mathematics and Physics. This is borne out by the relevance of these models in
such disparate fields as group theory~\cite{FGP05,SV04}, the theory of
special functions and orthogonal polynomials~\cite{BF97,Du98,Ta03,Ta05}, soliton
theory~\cite{Po95}, random matrix theory~\cite{SLA94,St04,TSA95}, quantum field
theory~\cite{CL99,DP98,SSAFR05}, etc. The Calogero and Sutherland models describe a system of
$N$ quantum particles in a line or circle, respectively, with pairwise interactions
inversely proportional to the square of the distance. Over the years, many different
generalizations of these models have been considered in the literature. One such significant
extension was proposed in the early eighties by Olshanetsky and Perelomov~\cite{OP83},
who showed that both the Calogero and Sutherland models are limiting cases of a more general integrable model
with a two-body interaction potential of elliptic type. The integrability of the
latter model was explained by expressing the Hamiltonian as one of the radial components
of the Laplace--Beltrami operator in a symmetric space associated with the $A_{N-1}$ root system.
It was also shown in Ref.~\cite{OP83} that one can construct integrable
generalizations of the Calogero--Sutherland (CS) models associated with any classical
(extended) root system, like $BC_N$.

Another essential feature of the original Calogero and Sutherland
models and their generalization to other root systems is their
exact solvability, that is, the fact that the whole spectrum can
be computed in closed form using algebraic techniques. In the last
decade, some authors have introduced further extensions of CS
models which are quasi-exactly solvable, in the sense that
only part of the spectrum can be computed
algebraically~\cite{GGR00,GGR01,HS99,MRT96}. In all of these
quasi-exactly solvable CS models, the Hamiltonian can be expressed
as a polynomial in the generators of a realization of the Lie
algebra $\fsl_{N+1}$ in terms of first-order differential
operators. Since these operators leave invariant a
finite-dimensional space of functions, the Hamiltonian is
guaranteed to possess a finite number of eigenfunctions belonging
to this space.

A great deal of attention has also been devoted to constructing
models of CS type for particles with internal degrees of freedom
(typically spin), partly motivated by their intimate connection
with integrable spin chains of Haldane--Shastry
type~\cite{Ha88,Sh88}. Two main approaches have been followed in
order to incorporate spin into CS models, based either on
supersymmetry~\cite{FM90,BTW98,DLM01,DLM03,Gh04} or the exchange (also known as Dunkl)
operator formalism~\cite{Ba96,Du89,Du98,FGGRZ03,Po92,Ya95}. The
spin models thus obtained include the exactly solvable spin
counterparts of the scalar Calogero and Sutherland models of $A_N$
and $BC_N$ type, as well as several quasi-exactly solvable
deformations thereof, some of them with elliptic potentials~\cite{FGGRZ01,FGGRZ01b}. A
common property shared by all of these models is the long-range
character of the interaction potential, in the sense that all
particles interact with each other.

The connection between spin CS models and spin chains of Haldane--Shastry
type was first elucidated by Polychronakos through a mechanism known as the ``freezing
trick''~\cite{Po93}. The main idea is that in the large coupling constant limit
the particles in a (dynamical) spin CS model freeze at the classical equilibrium
of the scalar part of the potential, thus giving rise to a spin chain with long-range
position-dependent interactions. In this limit the eigenfunctions of the spin CS model
factorize into the product of an eigenfunction of the corresponding scalar CS model times
an eigenfunction of the associated spin chain. If \emph{all} the eigenfunctions of both
the scalar and spin CS models are known, the partition function of the corresponding
spin chain can be exactly computed from those of the scalar and spin CS models~\cite{Po94,EFGR05,FG05}.

A few years ago, Jain and Khare presented a novel class of scalar
CS-like models of $A_N$ type, characterized by the fact that each
particle only interacts with its nearest and next-to-nearest
neighbors~\cite{JK99}. In a subsequent paper~\cite{AJK01}, Auberson, Jain and Khare
discussed a generalization of these models to the $BC_N$ root system and to higher dimensions.
The latter papers, however, left open some important issues, such as
the exact or quasi-exact solvability of these models, the
derivation of general explicit formulas for their eigenfunctions and
the existence of similar models for particles with spin. The
last question was first addressed by Deguchi and
Ghosh~\cite{DG01}, who introduced and partially solved several
spin $1/2$ extensions of the scalar models of Jain and Khare using
the supersymmetric approach. By a suitable generalization of the
exchange operator formalism, in our previous paper~\cite{EFGR05b}
we constructed the three spin models of $A_N$ type with near
neighbors interactions listed in Eq.~\eqref{Vs} below.
A significant property of these models is the fact that the spin
chains obtained from them by the freezing trick feature
short-range position-dependent interactions, and thus occupy an
intermediate position between the Heisenberg chain
(with short-range position-independent interactions) and the spin chains of Haldane--Shastry
type (possessing long-range position-dependent interactions).
In Ref.~\cite{EFGR05b} we presented without proof closed-form expressions
for several infinite families of
eigenfunctions of the scalar reductions of all three models,
considerably generalizing the results of Ref.~\cite{AJK01}. We
were also able to derive similar expressions for a wide class of spin
eigenfunctions of the models~\eqref{V1} and~\eqref{V2}. The
computation of the spin eigenfunctions for the remaining model~\eqref{V0},
which is probably the most interesting one due to the rich
structure of its finite-dimensional invariant spaces, was not
undertaken in Ref.~\cite{EFGR05b}.

In this paper we present a detailed analysis of the models~\eqref{Vs},
with special emphasis on the rational model~\eqref{V0}.
In particular, we have achieved a complete description of the
flag of invariant finite-dimensional spaces for the latter model
presented in Ref.~\cite{EFGR05b}. More importantly, we have found that
this flag can be further enlarged with an additional family of spin functions.
We have computed all the eigenfunctions of the model~\eqref{V0}
belonging to the new flag, thereby obtaining seven infinite
families of spin eigenfunctions in closed form. These eigenfunctions have been expressed in all
cases in a compact way in terms of generalized Laguerre and Jacobi polynomials.
The resulting expressions will be used in a forthcoming paper for computing a number of
eigenvalues and eigenfunctions of the spin chain obtained from the model~\eqref{V0}
by taking the strong coupling limit.

The paper is organized as follows. In Section~\ref{sec.mods} we define the
Hamiltonians of the spin many-body models which are the subject of this work,
and show that they can be expressed in terms of suitable differential operators
with near-neighbors exchange terms. Section~\ref{sec.spaces} is entirely devoted to
the characterization of certain finite-dimensional spaces of polynomial spin functions
invariant under these operators. The first part of this section deals with
the construction of the latter spaces and the proof of their invariance,
cf.~Theorem~\ref{thm.1} and Corollary~\ref{cor.1}. In the rest of the section
we complete the description of the invariant spaces for the model~\eqref{V0},
by identifying the spin states which satisfy a restriction stated in
Theorem~\ref{thm.1}. In Section~\ref{sec.eig} we
show that the eigenvalue problems for the Hamiltonians of the models~\eqref{Vs}
restricted to their invariant spaces reduce to finding the polynomial solutions
of a corresponding system of differential equations. By completely solving the latter problem,
we obtain several (infinite) families of eigenfunctions for the models~\eqref{Vs},
whose explicit expressions are presented in Theorems~\ref{thm.H0}--\ref{thm.H2}.
Finally, in Section~\ref{sec.summ} we summarize our results and outline some related
open problems.

\section{The models}\label{sec.mods}

In this section we shall introduce the three types of $N$-body models with
near-neighbors interactions whose study is the aim of this paper. We shall
also recall from our previous paper~\cite{EFGR05b} the relation of each
of these models with a corresponding differential-difference operator involving
near-neighbors exchange operators.

Let us begin with some preliminary definitions. We shall denote by
$\ket{s_1\dots s_N}$, where $s_i=-M,-M+1,\dots,M$ and
$M\in\frac12\NN$, the elements of the canonical basis of the space $\Si$ of
the particles' internal degrees of freedom ($\mathrm{SU}(2M+1)$
spin). The action of the spin permutation operators $S_{ij}$ on
this basis is given by
\[
S_{ij}\ket{\dots s_i\dots s_j\dots}=\ket{\dots s_j\dots s_i\dots}.
\]
The operators $S_{ij}$ can be expressed in terms of the fundamental
$\mathrm{SU}(2M+1)$ generators $S_i^a$, $a=1,\dots,4M(M+1)$, as
$S_{ij}=1/(2M+1)+\sum_a S_i^aS_j^a$. The Hamiltonians of the models we shall be
concerned with are given by
\begin{equation}\label{Hep}
H_\ep=-\sum_i\pa_{x_i}^2+V_\ep\,,\qquad\ep=0,1,2,
\end{equation}
where
\begin{subequations}\label{Vs}
\begin{align}
&\hspace*{-0.1em}V_0=\omega^2 r^2+\sum_i\frac{2a^2}{(x_i-x_{i-1})(x_i-x_{i+1})}+
\sum_i\frac{2a}{(x_i-x_{i+1})^2}\,(a-S_{i,i+1}),\label{V0}\\[1mm]
&\hspace*{-0.1em}V_1=\omega^2 r^2+\sum_i\frac{b(b-1)}{x_i^2}
+\sum_i\frac{8a^2x_i^2}{(x_i^2-x_{i-1}^2)(x_i^2-x_{i+1}^2)}\notag\\
&\hspace{15em}{}+4a\sum_i\frac{x_i^2+x_{i+1}^2}{{(x_i^2-x_{i+1}^2)}^2}
\,(a-S_{i,i+1}),\label{V1}\\[1mm]
&\hspace*{-0.1em}V_2=2a^2\sum_i\cot(x_i-x_{i-1})\cot(x_i-x_{i+1})
+2a\sum_i\csc^{2}(x_i-x_{i+1})\,(a-S_{i,i+1}),\label{V2}
\end{align}
\end{subequations}
with $r^2=\sum_i x_i^2$ and $a,b>1/2$. Here and in what follows,
all sums and products run from $1$ to $N$ unless otherwise stated, with the
identifications $x_0\equiv x_N$ and $x_{N+1}\equiv x_1$. A few
remarks on the configuration spaces of these models are now in
order. In all three models the potential diverges as
$(x_i-x_{i+1})^{-2}$ on the hyperplanes $x_i=x_{i+1}$,
so that the particles $i$ and $i+1$ cannot overtake one another.
Since we are interested in models with nearest and next-to-nearest
neighbors interactions, we shall henceforth assume that $x_1<\cdots<x_N$.
For the second potential~\eqref{V1} we shall take in addition
$x_1>0$, due to the double pole at $x_i=0$. For a similar reason,
we shall assume that $x_{i+1}-x_i<\pi$ for the potential~\eqref{V2}.

\begin{remark}
The Hamiltonians~\eqref{Hep} admit \emph{scalar reductions}
$H^{\mathrm{sc}}_\ep\equiv H_\ep\big|_{S_{i,i+1}\to 1}$ satisfying
the obvious identity
\[
H_\ep(\psi\ket s)=(H^{\mathrm{sc}}_\ep\psi)\ket s\,,
\]
where $\psi$ is a scalar function of the coordinates
$\bx=(x_1,\dots,x_N)$ and $\ket s$ is a totally symmetric spin
state. It follows that the spin Hamiltonians $H_\ep$ possess
\emph{factorized eigenfunctions} of the form $\Psi=\psi\ket s$,
where $\psi$ is an eigenfunction of the corresponding scalar
Hamiltonian $H^{\mathrm{sc}}_\ep$ and $\ket s$ is again a
symmetric spin state. The scalar
reductions of the models~\eqref{V0} and~\eqref{V2} were introduced
by Auberson, Jain and Khare in Ref.~\cite{AJK01}, whereas that of the model~\eqref{V1}
first appeared in our paper~\cite{EFGR05b}. It should also be noted
that for spin $1/2$ the potentials $V_0$ and $V_2$ differ from those studied by
Deguchi and Ghosh in Ref.~\cite{DG01} by a spin-dependent term.
\end{remark}

The models~\eqref{Vs} share a common
property that is ultimately responsible for their partial
solvability, namely that each Hamiltonian $H_\ep$ is related to a
scalar differential-difference operator involving near-neighbors exchange operators. Indeed, let
$K_{ij}$ denote the operator whose action on a smooth function
$f$ of the (possibly complex) coordinates $\bz=(z_1,\dots,z_N)$ is given by
\begin{equation}\label{KsAction}
(K_{ij}f)\,(z_1,\dots, z_i,\dots, z_j,\dots,z_N) =
f(z_1,\dots, z_j,\dots, z_i,\dots,z_N).
\end{equation}
Given a scalar differential-difference operator $D$ linear in the exchange operator
$K_{ij}$, we shall denote by
$D^*$ the differential operator acting on $C^{\infty}\otimes\Si$
obtained from $D$ by the replacement $K_{ij}\to S_{ij}$. One of the key ingredients
in our construction is the fact that
\begin{equation}\label{star}
D\ms\Phi=D^*\Phi\,,\qquad\text{for all}\quad \Phi\in\La\big(C^{\infty}\otimes\Si\big)\,,
\end{equation}
where $\La$ denotes the projector on states totally symmetric under simultaneous
permutations of the coordinates and spins. Consider next the
second-order differential-difference operators $T_\ep$ given by
\begin{equation}\label{Tep}
T_\ep=\sum_iz_i^\ep\pa_i^2+2a\sum_i\frac1{z_i-z_{i+1}}\,(z_i^\ep\pa_i-z_{i+1}^\ep\pa_{i+1})
-2a\sum_i\frac{\vartheta_\ep(z_i,z_{i+1})}{(z_i-z_{i+1})^2}\,(1-K_{i,i+1}),
\end{equation}
where $\pa_i=\pa_{z_i}$, $z_{N+1}\equiv z_1$, and
\[
\vartheta_0(x,y)=1\,,\qquad
\vartheta_1(x,y)=\frac12\,(x+y)\,,\qquad
\vartheta_2(x,y)=xy\,.
\]
Each Hamiltonian $H_\ep$ is related to a linear combination
\begin{equation}\label{BHep}
\BH_{\!\ep}=c\,T_\ep+c_- J^-+c_0 J^0+E_0
\end{equation}
of its corresponding operator $T_\ep$ and the first-order differential
operators
\begin{equation}\label{J-J0}
J^-=\sum_i \pa_i\,,\qquad J^0=\sum_i z_i\pa_i
\end{equation}
through the star mapping, a change of variables and a gauge transformation.
More precisely,
\begin{equation}\label{Hepstar}
H_\ep=\mu\cdot\BH_{\!\ep}^*\big|_{z_i=\ze(x_i)}\cdot\mu^{-1}\,,\qquad \ep=0,1,2,
\end{equation}
where the constants $c$, $c_-$, $c_0$, $E_0$, the gauge factor $\mu(\bx)$, and the
change of variables $\ze(x)$ for each model are listed in Table~\ref{table:params}.

\begin{table}[h]
\caption{Parameters, gauge factor and change of variable in Eqs.~\eqref{BHep}
and~\eqref{Hepstar}.}\label{table:params}
\begin{center}
\begin{tabular}{lccc}\hline
\BStrut & $\ep=0$ & $\ep=1$ & $\ep=2$\\ \hline
\BStrut $c$ & $-1$ & $-4$ & $4$\\ \hline
\BStrut $c_-$ & $0$ & $-2(2b+1)$ & $0$\\ \hline
\BStrut $c_0$ & $2\om$ & $4\om$ & $4(1-2a)$\\ \hline
\BStrut $E_0$ & $N\omega(2a+1)$ & $N\omega(4a+2b+1)$ & $2Na^2$\\ \hline
\BStrut $\mu(\bx)\quad$ & $\;\e^{-\frac\omega2\,r^2}\prod\limits_i|x_i-x_{i+1}|^a\;$
& $\;\e^{-\frac\omega2\,r^2}\prod\limits_i{|x_i^2-x_{i+1}^2|}^a\,x_i^b\;$
& $\;\prod\limits_i\sin^a|x_i-x_{i+1}|\;$\\ \hline
\BStrut $\ze(x)$ & $x$ & $x^2$ & $\e^{\pm2\iu x}$\\ \hline
\end{tabular}
\end{center}
\end{table}

{}From Eqs.~\eqref{star} and~\eqref{Hepstar} it follows that if
$\Phi(\bz)\in\La\big(C^{\infty}\otimes\Si\big)$ is a symmetric eigenfunction of
$\BH_{\!\ep}$, then
\begin{equation}\label{PsimuPhi}
\Psi(\bx)=\mu(\bx)\Phi(\bz)|_{z_i=\ze(x_i)}
\end{equation}
is a (formal) eigenfunction of $H_{\ep}$ with the same eigenvalue. In this paper we shall
construct a flag $\cH^{\ms 0}_\ep\subset\cH^{\ms 1}_\ep\subset\cdots$ of
finite-dimensional subspaces of $\La\big(\CC[\bz]\otimes\Si\big)$ invariant under
each $\BH_{\!\ep}$. We will show that the problem of
diagonalizing $\BH_{\!\ep}$ in each subspace $\cH^{\ms n}_\ep$
is equivalent to the computation of the polynomial solutions
of a system of linear differential equations. We shall completely
solve this problem, thereby obtaining several
infinite families of eigenfunctions of~$H_{\!\ep}$ for each $\ep$.
{}From the expressions for the change of variable and the gauge factor in
Table~\ref{table:params}, and the fact that the functions $\Phi$ in
Eq.~\eqref{PsimuPhi} are in all cases polynomials,
it immediately follows that the eigenfunctions thus obtained
are in fact normalizable.

\begin{remark}
The operators \eqref{Tep} can be expressed as quadratic combinations of the
first-order operators
\[
D_i^\ep = z^\ep_i\pa_i\,,\qquad
Q_i^\ep = \frac{\vartheta_\ep(z_i,z_{i+1})}{z_i-z_{i+1}}\,(1-K_{i,i+1})
+\frac{\vartheta_\ep(z_{i-1},z_i)}{z_i-z_{i-1}}\,(1-K_{i-1,i})\,,
\]
where $\ep=0,1$, as follows:
\begin{align*}
& T_0=\sum_i\Big[{\big(D_i^0\big)}^2+a\big\{D_i^0,Q_i^0\big\}\Big]\,,\qquad
T_1=\sum_i\Big[D_i^1D_i^0+a\big\{D_i^1,Q_i^0\big\}\Big]\,,\\
& T_2=\sum_i\Big[{\big(D_i^1\big)}^2+\Big\{D_i^1,a\ms Q_i^1+a-\frac12\Big\}\Big]\,.
\end{align*}
For each nonnegative integer $n$, the space $\cP^n$ of polynomials in
$\bz$ of total degree at most $n$ is invariant under the operators
$Q^\ep_i$ (see Ref.~\cite{FGGRZ01}), and hence also under both $T_\ep$ and $\BH_{\!\ep}$.
Note, however, that the operators $\BH_{\!\ep}$ do not commute with
the symmetrizer $\La$, and thus the previous observation does not imply that
they preserve the space $\La(\cP^n\otimes\Si)$ of symmetric spin functions of
polynomial type. Consequently, $\BH_{\!\ep}$ is not guaranteed \emph{a priori}
to admit finite-dimensional invariant subspaces of
$\La\big(\CC[\bz]\otimes\Si\big)$. This is in fact the main difference with the
usual solvable spin CS
models~\cite{HH92,MP93,BGHP93,Ba96,Du98,FGGRZ01,FGGRZ01b}, for which the
operators analogous to $\BH_{\!\ep}$ preserve $\cP_n$ and commute with $\La$,
and hence automatically leave invariant the space $\La(\cP^n\otimes\Si)$.
\end{remark}

\section{The invariant spaces}\label{sec.spaces}

In this section we shall prove that each operator $T_\ep$
leaves invariant a flag $\cT^{\ms 0}_\ep\subset\cT^{\ms 1}_\ep\subset\cdots$,
where $\cT^{\ms n}_\ep$ is a finite-dimensional subspace of $\La(\cP^n\otimes\Si)$.
This result will then be used to construct
a corresponding invariant flag \mbox{$\cH^{\ms 0}_\ep\subset\cH^{\ms 1}_\ep\subset\cdots$}
for the operator $\BH_{\!\ep}$, where $\cH^{\ms n}_\ep\subset\cT^{\ms n}_\ep$ for all $n$.

Let us first introduce the following two sets of elementary symmetric
polynomials:
\[
\si_k=\sum_i z_i^k\,,\qquad \tau_k=\sum\limits_{i_1<\cdots<i_k}z_{i_1}\cdots z_{i_k}\,;
\qquad k=1,\dots,N\,.
\]
It is well known that any symmetric polynomial in $\bz$ can be expressed
as a polynomial in either $\bsi\equiv(\si_1,\dots,\si_N)$ or
$\btau\equiv(\tau_1,\dots,\tau_N)$.

We shall denote by $2aX_\ep$ the terms of $T_\ep$ linear in derivatives, that is
\[
X_\ep=\sum_i\frac1{z_i-z_{i+1}}\,(z_i^\ep\pa_i-z_{i+1}^\ep\pa_{i+1})\,.
\]
In the next lemma we show that each vector field $X_\ep$ leaves invariant a
corresponding flag $\cX_\ep^0\subset\cX_\ep^1\subset\cdots$ of finite-dimensional
subspaces of the space $\cS\equiv\CC[\bsi]=\CC[\btau]$ of symmetric polynomials in $\bz$.
\begin{lemma}
For each $n=0,1,\dots$, the operator $X_\ep$ leaves invariant the linear space $\cX_\ep^n$, where
\[
\cX_0^n=\CC[\si_1,\si_2,\si_3]\cap\cP^n,\quad
\cX_1^n=\CC[\si_1,\si_2,\tau_N]\cap\cP^n,\quad
\cX_2^n=\CC[\si_1,\tau_{N-1},\tau_N]\cap\cP^n.
\]
\end{lemma}
\begin{proof}
If $f$ is a function of the symmetric variables $\si_1,\si_2,\si_3,\tau_{N-1},\tau_N$,
we shall use from now on the convenient notation
\[
f_k=\begin{cases}
\pa_{\si_k}f\,,\quad & k=1,2,3,\\[1mm]
\pa_{\tau_k}f\,,\quad & k=N-1,N.
\end{cases}
\]
Let us first consider the vector field $X_0$. Since
\[
X_0\si_k=k\sum_iz_i^{k-1}X_0z_i=k\bigg(\sum_i\frac{z_i^{k-1}}{z_i-z_{i+1}}
-\sum_i\frac{z_i^{k-1}}{z_{i-1}-z_i}\bigg)=
\begin{cases}
0\,, & k=1\,,\\
2N\,, & k=2\,,\\
6\si_1\,, & k=3\,,
\end{cases}
\]
if $f\in\cX^n_0$ we have
\begin{subequations}\label{Xf}
\begin{equation}\label{X0f}
X_0f=2(Nf_2+3\si_1 f_3)\in\cX^n_0\,.
\end{equation}
The proof for the remaining two cases follows from the analogous formulas
\begin{align}
X_1f&=Nf_1+4\si_1f_2\,,& f\in\cX^n_1\,;\label{X1f}\\
X_2f&=2\si_1f_1+N(\tau_{N-1}f_{N-1}+\tau_Nf_N)\,,& f\in\cX^n_2\,.\label{X2f}
\end{align}
\end{subequations}
\end{proof}
\begin{remark}
It should be noted that these flags cannot be trivially enlarged,
since, e.g.,
\begin{align*}
\frac14\,X_0\si_4&=2\si_2+\sum_iz_iz_{i+1}\,,\\
\frac13\,X_1\si_3&=2\si_2+\sum_iz_iz_{i+1}\,,&
X_1\tau_{N-1}&=\tau_N\sum_i({z_iz_{i+1}})^{-1}\,,\\
\frac12\,X_2\si_2&=2\si_2+\sum_iz_iz_{i+1}\,,&
X_2\tau_{N-2}&=N\tau_{N-2}-\tau_N\sum_i(z_iz_{i+1})^{-1}
\end{align*}
are not symmetric polynomials.
\end{remark}
We note that the restriction of $T_\ep$ to $\cX^n_\ep\subset\cS$ obviously satisfies
\begin{equation}\label{Tep-res}
T_\ep|_{\cX^n_\ep}=\sum_iz_i^\ep\pa_i^2+2aX_\ep\,.
\end{equation}
The second-order terms of the operator~\eqref{Tep-res}, however, do not preserve
the corresponding space $\cX^n_\ep$, unless one imposes the
additional restrictions specified in the following proposition:
\begin{proposition}\label{prop.cS}
For each $n=0,1,\dots$, the operator $T_\ep$ leaves invariant the linear space $\cS_\ep^n$, where
\begin{align*}
\cS_0^n&=\{f\in\cX_0^n\mid f_{33}=0\}\,,\\
\cS_1^n&=\{f\in\cX_1^n\mid f_{22}=f_{NN}=0\}\,,\\
\cS_2^n&=\{f\in\cX_2^n\mid f_{11}=f_{N-1,N-1}=0\}\,.
\end{align*}
\end{proposition}
\begin{proof}
Let us begin with the operator $T_0$. If $f\in\cX_0^n$,
an elementary computation shows that
\begin{equation}\label{paif0}
\pa_if=f_1+2z_if_2+3z_i^2f_3\\
\end{equation}
and therefore
\begin{multline}\label{T20f}
\sum_i \pa_i^2f=N(f_{11}+2f_2)+
2(2f_{12}+3f_3)\si_1\\
{}+2(3f_{13}+2f_{22})\si_2
+12f_{23}\si_3+9f_{33}\si_4\,.
\end{multline}
{}From the previous formula and Eq.~\eqref{X0f} it follows that $T_0f\in\cS_0^n$
whenever $f\in\cS_0^n$. Similarly, if $f\in\cX^n_1$ we have
\begin{equation}\label{paif1}
  \pa_if=f_1+2z_if_2+z_i^{-1}{\tau_N}f_N\,,
\end{equation}
so that
\begin{multline}\label{T21f}
\sum_i z_i\pa_i^2f=(f_{11}+2f_2)\si_1+4f_{12}\si_2+4f_{22}\si_3\\
{}+2Nf_{1N}\tau_N+4f_{2N}\si_1\tau_N+f_{NN}\tau_{N-1}\tau_N\,,
\end{multline}
which together with Eq.~\eqref{X1f} implies that $T_1f\in\cS^n_1$ for all $f\in\cS^n_1$.
Finally, if $f\in\cX^n_2$ then
\begin{equation}\label{paif2}
\pa_if=f_1+\big(z_i^{-1}{\tau_{N-1}}-{z_i^{-2}}{\tau_N}\big)f_{N-1}+z_i^{-1}{\tau_N}f_N
\end{equation}
and hence
\begin{multline}\label{T22f}
  \sum_i
  z_i^2\pa_i^2f=f_{11}\si_2+2f_{1,N-1}(\si_1\tau_{N-1}-N\tau_N)\\[-1mm]
  +2f_{1N}\si_1\tau_N+f_{N-1,N-1}\big[(N-1)\tau_{N-1}^2-2\tau_{N-2}\tau_N\big]\\[2mm]
  +2(N-1)f_{N-1,N}\tau_{N-1}\tau_N+Nf_{NN}\tau_N^2\,.
\end{multline}
The statement follows again from the previous equation and Eq.~\eqref{X2f}.
\end{proof}
The last proposition implies that each operator $T_\ep$
preserves ``trivial'' symmetric spaces $\cS^n_\ep\otimes\La(\Si)$ spanned by
factorized states. The main theorem of this section shows that in fact the
latter operator leaves invariant a flag of nontrivial finite-dimensional
subspaces of $\La(\cP^n\otimes\Si)$. Before stating this theorem we need to
make a few preliminary definitions. Given a spin state $\ket s\in\Si$, we set
\begin{equation}\label{sisij}
\ket{s_i}=\frac1{N!}\sum_{\substack{\pi\in S_N\\\pi(1)=i}}\pi\ms\ket s\,,\qquad\quad
\ket{s_{ij}^\pm}=\frac1{N!}\!\!
\sum_{\substack{\pi\in S_N\\\pi(1)=i,\pi(2)=j}}\!\!\pi\ms(1\pm S_{12})\ms\ket s\,,
\end{equation}
where $S_N$ is the symmetric group on $N$ elements. Here and throughout the paper
we identify an abstract permutation $\pi$ with its realization as a permutation of the
particles' spins. {}From Eq.~\eqref{sisij} we have
\begin{equation}\label{Lafs}
\La\big(f(z_1)\ket s\big)=\sum_if(z_i)\ket{s_i}\,,\qquad
\La\big(g^\pm(z_1,z_2)\ket s\big)=\sum_{i<j}g^\pm(z_i,z_j)\ket{s^\pm_{ij}}\,,
\end{equation}
where the last identity holds if $g^\pm(z_2,z_1)=\pm g^\pm(z_1,z_2)$. We also define the subspace
\begin{equation}\label{Sip}
\Si'=\Big\{\,\ket s\in\Si\;\big|
\;\,{\textstyle\sum\limits_i}\ket{s^+_{i,i+1}}\in\La(\Si)\,\Big\}\subset\Si\,.
\end{equation}
\begin{theorem}\label{thm.1}
Let
\begin{align*}
\cT^n_0 &=\big\langle f(\si_1,\si_2,\si_3)\La\ket s,g(\si_1,\si_2,\si_3)\La(z_1\ket s),
h(\si_1,\si_2)\La(z_1^2\ket s),\\
&\hspace{4em}\Th(\si_1,\si_2)\La(z_1z_2\ket{s'}),w(\si_1,\si_2)\La(z_1z_2(z_1-z_2)\ket{s})
\mid f_{33}=g_{33}=0\big\rangle\,,\\[1mm]
\cT^n_1 &=\big\langle f(\si_1,\si_2,\tau_N)\La\ket s,
g(\si_1,\tau_N)\La(z_1\ket s)
\mid{}f_{22}=f_{NN}=g_{NN}=0\big\rangle\,,\\[1mm]
\cT^n_2&=\big\langle f(\si_1,\tau_{N-1},\tau_N)\La\ket s,g(\tau_{N-1},\tau_N)\La(z_1\ket s),
\tau_Nq(\si_1,\tau_N)\La(z_1^{-1}\ket s)\\
&\hspace{14.2em}{}\mid f_{11}=f_{N-1,N-1}=g_{N-1,N-1}=q_{11}=0\big\rangle\,,
\vrule depth6pt width0pt
\end{align*}
where $\ket s\in\Si$, $\ket{s'}\in\Si'$, $\deg f\leq n$, $\deg g\leq n-1$,
$\deg h\leq n-2$, $\deg\Th\leq n-2$, $\deg w\leq n-3$, $\deg q\leq n-N+1$,
and $\deg$ is the total degree in $\bz$. Then $\cT_\ep^n$ is invariant under
$T_\ep$ for all $n=0,1,\dots$.
\end{theorem}
\begin{proof}
By Proposition~\ref{prop.cS}, it suffices to show that $T_\ep$ maps
$\cT_\ep^n/\big(\cS^n_\ep\otimes\La(\Si)\big)$ into $\cT_\ep^n$.
We shall first deal with the operator $T_0$. Consider the states
of the form $g\La(z_1\ket s)$, with $g\in\cS_0^{n-1}$. Since
\[
(\pa_l-\pa_{l+1})z_i=\frac1{z_l-z_{l+1}}\,(1-K_{l,l+1})z_i\,,\qquad\forall\:i,l,
\]
we have
\[
T_0(gz_i)=(T_0g)z_i+2\pa_ig\,.
\]
Calling
\begin{equation}\label{Phik}
\Phi^{(k)}\equiv \La(z_1^k\ket s),\qquad k\in\ZZ\,,
\end{equation}
from Eqs.~\eqref{paif0} and~\eqref{Lafs} we obtain
\begin{equation}\label{Phi01}
T_0\big(g\Phi^{(1)}\big)=\sum_iT_0(gz_i)\ket{s_i}=
(T_0g)\Phi^{(1)}+2\sum_{k=1}^3kg_k\Phi^{(k-1)}\in\cT_0^{n-2}\,.
\end{equation}
Similarly, if $h(\si_1,\si_2)\in\cS_0^{n-2}$, the identity
\[
(\pa_l-\pa_{l+1})z_i^2=\frac1{z_l-z_{l+1}}\,(1-K_{l,l+1})z_i^2+(z_l-z_{l+1})
(\de_{li}+\de_{l,i-1})\,,\qquad\forall\:i,l
\]
implies that
\[
T_0(hz_i^2)=(T_0h)z_i^2+4z_i\pa_ih+2(2a+1)h\,,
\]
and therefore
\begin{align}
T_0\big(h\Phi^{(2)}\big)&=\sum_iT_0(hz_i^2)\ket{s_i}\notag\\
&=(T_0h+8h_2)\Phi^{(2)}+4h_1\Phi^{(1)}+2(2a+1)h\Phi^{(0)}\label{Phi02}
\end{align}
belongs to $\cT_0^{n-2}$ on account of Eqs.~\eqref{X0f} and~\eqref{T20f}.
On the other hand, from the equality
\[
(\pa_l-\pa_{l+1})z_iz_j=\frac1{z_l-z_{l+1}}\ms(1-K_{l,l+1})z_iz_j-(z_l-z_{l+1})
\de_{j,i+1}\de_{l,i},\quad\forall\:i<j,\forall\:l
\]
it follows that
\[
T_0(\tilde hz_iz_j)=(T_0\Th)z_iz_j+2(z_i\pa_j\Th+z_j\pa_i\Th)-2a\Th\de_{j,i+1}\,.
\]
Setting
\begin{equation}\label{tPhi2}
\tPhi^{(2)}\equiv\La(z_1z_2\ket s)
\end{equation}
and using again Eqs.~\eqref{paif0} and~\eqref{Lafs} we then have
\begin{align}
T_0\big(\Th\tPhi^{(2)}\big)&=\sum_{i<j}T_0(\tilde hz_iz_j)\ket{s^+_{ij}}\notag\\
&=(T_0\Th+8\Th_2)\tPhi^{(2)}+2\Th_1\La\big[(z_1+z_2)\ket s\big]
-2a\Th\sum_i\ket{s^+_{i,i+1}}\,.\label{tPhi02}
\end{align}
Since $\La\big[(z_1+z_2)\ket s\big]=\La\big[z_1(1+S_{12})\ket s\big]$,
the RHS of Eq.~\eqref{tPhi02} belongs to $\cT_0^{n-2}$ if and only if $\ket s\in\Si'$.
The last type of states generating the module $\cT_0^n$ are of the form $w(\si_1,\si_2)\hPhi^{(3)}$,
where
\begin{equation}\label{hPhi3}
\hPhi^{(3)}\equiv\La(z_1z_2(z_1-z_2)\ket s)\,.
\end{equation}
{}From the equality
\begin{multline*}
\frac1{z_l-z_{l+1}}\,(\pa_l-\pa_{l+1})\big[z_iz_j(z_i-z_j)\big]
=\frac1{(z_l-z_{l+1})^2}\,(1-K_{l,l+1})\big[z_iz_j(z_i-z_j)\big]
\\+(\de_{l,i-1}+\de_{li})z_j-(\de_{l,j-1}+\de_{lj})z_i,\qquad\forall\:i<j,\forall\:l
\end{multline*}
it follows that
\begin{multline}
T_0\big[w z_iz_j(z_i-z_j)\big]=(T_0w)z_iz_j(z_i-z_j)+2z_j(2z_i-z_j)\pa_iw\\
-2z_i(2z_j-z_i)\pa_jw-2(2a+1)(z_i-z_j)w\,.
\end{multline}
Using again Eqs.~\eqref{paif0} and~\eqref{Lafs} we obtain
\begin{multline}
T_0\big(w\hPhi^{(3)}\big)=\sum_{i<j}T_0\big(wz_iz_j(z_i-z_j)\big)\ket{s^-_{ij}}
=(T_0w+12w_2)\hPhi^{(3)}\\
+2w_1\La\big[(z_1^2-z_2^2)\ket s\big]
-2(2a+1)w\La\big[(z_1-z_2)\ket s\big]\,.\label{hPhi03}
\end{multline}
Since $\La\big[(z_1^k-z_2^k)\ket s\big]=\La\big[z_1^k(1-S_{12})\ket s\big]$,
the RHS of the latter equation clearly belongs to $\cT_0^{n-2}$.
This shows that $T_0(\cT_0^n)\subset\cT_0^{n-2}\subset\cT_0^n$.

Consider next the action of the operator $T_1$ on a state of the form
$g(\si_1,\tau_N)\Phi^{(1)}$, with $g\in\cS_1^{n-1}$. From the identity
\[
(z_l\pa_l-z_{l+1}\pa_{l+1})z_i=\frac12\,\frac{z_l+z_{l+1}}{z_l-z_{l+1}}\,(1-K_{l,l+1})z_i
+\frac12\,(z_l-z_{l+1})(\de_{l,i}+\de_{l,i-1})\,,\quad\forall\:i,l
\]
we easily obtain
\[
T_1(gz_i)=(T_1g)z_i+2z_i\pa_ig+2ag\,,
\]
and therefore, by Eqs.~\eqref{X1f}, \eqref{paif1} and~\eqref{T21f},
\begin{equation}\label{Phi11}
T_1\big(g\Phi^{(1)}\big)=\sum_i T_1(gz_i)\ket{s_i}=
(T_1g+2g_1)\Phi^{(1)}+2(ag+\tau_Ng_N)\Phi^{(0)}\in\cT_1^{n-1}\,.
\end{equation}
Thus $T_1(\cT_1^n)\subset\cT_1^{n-1}\subset\cT_1^n$, as claimed.

Consider, finally, the operator $T_2$. If $g(\tau_{N-1},\tau_N)\in\cS^{n-1}_2$,
the identity
\[
(z_l^2\pa_l-z_{l+1}^2\pa_{l+1})z_i=\frac{z_lz_{l+1}}{z_l-z_{l+1}}(1-K_{l,l+1})z_i
+z_i(z_l-z_{l+1})(\de_{l,i}+\de_{l,i-1}),\quad\forall\,i,l
\]
yields
\[
T_2(gz_i)=(T_2g)z_i+2z_i^2\pa_ig+4az_ig\,,
\]
and hence, by Eq.~\eqref{paif2},
\begin{align}
T_2(g\Phi^{(1)})&=\sum_i T_2(gz_i)\ket{s_i}\notag\\
&=\big[T_2g+2(\tau_{N-1}g_{N-1}+\tau_Ng_N+2ag)\big]\Phi^{(1)}
-2\tau_Ng_{N-1}\Phi^{(0)}\label{Phi21}
\end{align}
clearly belongs to $\cT_2^n$ on account of Eqs.~\eqref{X2f} and~\eqref{T22f}.
The last type of spin states we need to study are of the form
$\hq\Phi^{(-1)}$, where $\hq\equiv\tau_Nq(\si_1,\tau_N)$ with $q_{11}=0$. Since
\[
(z_l^2\pa_l-z_{l+1}^2\pa_{l+1})z_i^{-1}
=\frac{z_lz_{l+1}}{z_l-z_{l+1}}\,(1-K_{l,l+1})z_i^{-1}\,,\qquad\forall\:i,l,
\]
we obtain
\[
T_2\big(\hq z_i^{-1}\big)=(T_2\hq)z_i^{-1}-2\pa_i\hq+2\hq z_i^{-1}\,,
\]
and thus, by Eqs.~\eqref{paif2} and~\eqref{Lafs},
\begin{equation}\label{Phi2-1}
T_2\big(\hq\Phi^{(-1)}\big)=\sum_i T_2(\hq z_i^{-1})\ket{s_i}
=(T_2\hq-2\tau_N\hq_N+2\hq)\Phi^{(-1)}-2\hq_1\Phi^{(0)}\,.
\end{equation}
{}From Eqs.~\eqref{X2f} and~\eqref{T22f}, it follows that the RHS of the previous
equation belongs to $\cT_2^n$. Hence $T_2(\cT_2^n)\subset\cT_2^n$, which concludes the
proof.
\end{proof}
\begin{remark}
We have chosen to allow a certain
overlap between the different types of states spanning the spaces $\cT^n_\ep$.
For instance, if $\ket s$ is symmetric the state
$g(\si_1,\si_2,\si_3)\La(z_1\ket s)\in\cT_0^n$ is also of the form
$f(\si_1,\si_2,\si_3)\La\ket s$. Less trivially,
if $\ket s$ involves only two distinct spin components and is antisymmetric under $S_{12}$,
then we have
\[
\hPhi^{(3)}=\frac2N\big(\si_1\Phi^{(2)}-\si_2\Phi^{(1)}\big)\,,
\]
where $\Phi^{(k)}$ and $\hPhi^{(3)}$ are respectively defined in Eqs.~\eqref{Phik} and~\eqref{hPhi3}.
Hence, for spin $1/2$ the states of the form $w(\si_1,\si_2)\hPhi^{(3)}$ in the space
$\cT_0^n$ can be expressed in terms of the other generators of this space.
\end{remark}

The main result of this section follows easily from the previous theorem:

\begin{corollary}\label{cor.1}
For each $\ep=0,1,2$, the gauge Hamiltonian $\BH_\ep$ leaves invariant the space $\cH^n_\ep$
defined by
\begin{equation}\label{BcHs}
\cH^n_0=\cT^n_0\,,\qquad
\cH^n_1=\cT^n_1\big|_{f_N=g_N=0}\,,\qquad
\cH^n_2=\cT^n_2\,.
\end{equation}
\end{corollary}
\begin{proof}
We shall begin by showing that each space $\cT_\ep^n$ is invariant under
the operator $J^0$. Note first that
\begin{equation}\label{J0Phis}
J^0\Phi^{(j)}=j\ms\Phi^{(j)}\,,\qquad
J^0\tPhi^{(2)}=2\ms\tPhi^{(2)}\,,\qquad
J^0\hPhi^{(3)}=3\ms\hPhi^{(3)}\,;\qquad j\in\ZZ\,,
\end{equation}
where the states $\Phi^{(j)}$, $\tPhi^{(2)}$ and $\hPhi^{(3)}$ are
defined in Eqs.~\eqref{Phik}, \eqref{tPhi2}
and~\eqref{hPhi3}, respectively. Using Eqs.~\eqref{paif0}, \eqref{paif1} and~\eqref{paif2}
one can immediately establish the identities
\begin{subequations}\label{J0fs}
\begin{align}
J^0f&=\si_1f_1+2\si_2f_2+3\si_3f_3\,,\qquad & &\forall f(\si_1,\si_2,\si_3)\,,\label{J0f0}\\
J^0f&=\si_1f_1+2\si_2f_2+N\tau_Nf_N\,,\qquad & &\forall f(\si_1,\si_2,\tau_N)\,,\label{J0f1}\\
J^0f&=\si_1f_1+(N-1)\tau_{N-1}f_{N-1}+N\tau_Nf_N\,,\qquad & &\forall f(\si_1,\tau_{N-1},\tau_N)\,.
\label{J0f2}
\end{align}
\end{subequations}
{}From Eqs.~\eqref{J0Phis}-\eqref{J0fs} and the fact that $J^0$ is a derivation it
follows that $J^0$ leaves invariant the spaces $\cT^n_\ep$ for all $\ep=0,1,2$.
This implies that $\BH_\ep$ preserves $\cT^n_\ep$ for $\ep=0,2$, since the coefficient
$c_-$ vanishes in these cases (cf.~Table~\ref{table:params}). On the other hand,
for $\ep=1$ the coefficient $c_-$ is nonzero, and thus we have to consider the action
of the operator $J^-$ on the space $\cT^n_1$. We now have
\begin{equation}\label{JmPhis}
J^-\Phi^{(j)}=j\ms\Phi^{(j-1)}\,,\qquad j\in\ZZ\,,
\end{equation}
and, from Eq.~\eqref{paif1},
\begin{equation}\label{Jmf1}
J^-f=Nf_1+2\si_1f_2+\tau_{N-1}f_N\,,\qquad\forall f(\si_1,\si_2,\tau_N)\,.
\end{equation}
Hence $J^-$ leaves invariant the subspace $\cH^n_1$ of $\cT^n_1$
defined by the restrictions $f_N=g_N=0$. {}From the obvious identity
$T_1\big(f\Phi^{(0)}\big)=\big(T_1f\big)\Phi^{(0)}$
and Eq.~\eqref{Phi11}, together with~\eqref{X1f}, \eqref{Tep-res}
and~\eqref{T21f}, it follows that the operator $T_1$ also preserves~$\cH^n_1$.
Likewise, Eqs.~\eqref{J0Phis} and~\eqref{J0f1} imply that $\cH^n_1$
is invariant under $J^0$, and hence under the gauge Hamiltonian $\BH_1$.
\end{proof}

Theorem~\ref{thm.1} characterizes the invariant space $\cT_0^n$
in terms of the subspace $\Si'\subset\Si$ in Eq.~\eqref{Sip} that
we shall now study in detail. In fact, from the definition of the
invariant space $\cT_0^n$ it follows that we can consider without
loss of generality the quotient space $\Si'/{\sim}$, where $\ket
s\sim\ket{\ts}$ if $\La(z_1z_2\ket{s})=\La(z_1z_2\ket{\ts})$. For
instance, from Eq.~\eqref{sisij} it immediately follows that if
$\ket{s}\in\Si'$ and $\pi\in S_N$ is a permutation such that
$\pi(i)\in\{1,2\}$ for $i=1,2$, then $\pi\ket{s}$ belongs to
$\Si'$ and is equivalent to $\ket s$.

In the rest of this section, we shall denote $\ket{s^+_{ij}}$
simply as $\ket{s_{ij}}$ for the sake of conciseness. From
Eq.~\eqref{sisij} it easily follows that any symmetric state belongs to $\Si'$, since
\begin{equation}\label{AsLas}
\sum_i\ket{s_{i,i+1}}=\frac2{N-1}\,\ket s\,,\quad\text{for all }\ket s\in\La(\Si)\,.
\end{equation}
On the other hand, if $\ket s\in\La(\Si)$ the corresponding state
$h(\si_1,\si_2)\La(z_1z_2\ket s)$ is a trivial (factorized) state.
We shall next show that the reciprocal of this statement is also true,
up to equivalence.
\begin{lemma}
For every $\ket s\in\Si$, $\La(z_1z_2\ket{s})$ is a factorized state if and only if
$\ket s\sim\La\ket s$.
\end{lemma}
\begin{proof}
Suppose that
\[
\La(z_1z_2\ket{s})=\ket\hs\sum_{i<j}c_{ij}z_iz_j
\]
is a factorized state. Since the LHS of the previous formula is symmetric,
 $c_{ij}=c$ for all $i,j$ and $\ket\hs\in\La(\Si)$. By absorbing the constant
$c$ into $\ket\hs$ we can take $c=1$ without loss of generality, and
therefore
\[
\La(z_1z_2\ket{s})=\sum_{i<j}z_iz_j\ket{s_{ij}}=\tau_2\ket\hs
\quad\implies\quad \ket{s_{ij}}=\ket\hs\,,\qquad
i,j=1,\dots\,,N.
\]
From Eq.~\eqref{Lafs} with $f(z_1,z_2)=1$ it then follows that
\[
\La\ket s = \sum_{i<j}\ket{s_{ij}}=\frac12\,N(N-1)\ket\hs\,.
\]
Setting $\ket{s_0}=\ket s-\La\ket s$ and using the previous identity we
obtain
\[
  \La(z_1z_2\ket{s_0})=\La(z_1z_2\ket{s})-\frac{2\ms\tau_2}{N(N-1)}\,\La\ket s
  =\ket\hs\tau_2-\frac{2\ms\tau_2}{N(N-1)}\,\La\ket s=0\,.
\]
Hence $\ket s\sim\La\ket s$, as claimed.
\end{proof}
By the previous observations, it suffices to characterize the
nonsymmetric states in $\Si'$. To this end, let us introduce the
linear operator $A:\Si\to\Si$ by
\begin{equation}\label{A}
A\ket s=\sum_i\ket{s_{i,i+1}}.
\end{equation}
Given an element $\ket\bs\equiv\ket{s_1\dots s_N}$ of the
canonical basis of $\Si$, we shall also denote by
$\{s^1,\dots,s^n\}$ the set of distinct components of
$\bs\equiv(s_1,\dots,s_N)$, and by $\nu_i$ the number of times
that $s^i$ appears among the components of $\bs$. For instance, if
$\ket\bs=\ket{{-2},0,1,-2,1}$, then we can take $s^1=-2$, $s^2=0$,
$s^3=1$, so that $\nu_1=\nu_3=2$, $\nu_2=1$. Consider the spin
states $\ket{\chi_i(\bs)}\equiv\ket{\chi_i}$, $i=1,\dots,n$, given
by
\begin{subequations}\label{chis}
\begin{align}
\ket{\chi_i}&=
\nu_i(\nu_i-1)\ket{s^is^i\dots}-\sum_{\substack{1\leq j,k\leq n\\j,k\neq i}}\nu_j(\nu_k-\de_{jk})
\ket{s^js^k\dots}\,,\qquad &\nu_i>1\,,\label{chia}\\[1mm]
\ket{\chi_i}&=\sum_{\substack{1\leq j\leq n\\j\neq i}}\nu_j\,
\big(\ket{s^is^j\dots}+\ket{s^js^i\dots}\big)\,,\qquad &\nu_i=1\,.\label{chib}
\end{align}
\end{subequations}
Here we have adopted the following convention: an ellipsis inside a ket stands
for an arbitrary ordering of the components in $\bs$ not indicated
explicitly. Note that the states~\eqref{chis} are defined only up to
equivalence, and that $\ket{\chi_i(\bs)}=\ket{\chi_i(\pi\bs)}$ for any permutation
$\pi\in S_N$.
\begin{proposition}
Given a basic spin state $\ket\bs$, the associated spin states $\ket{\chi_i(\bs)}$
are all in $\Si'/{\sim}$.
\end{proposition}
\begin{proof}
Consider first a state $\ket{\chi_i}$ of the type~\eqref{chia}. Using the
definition of the operator $A$ in Eq.~\eqref{A} we obtain
\begin{multline}\label{Achia}
N!A\ket{\chi_i}=2\nu_i(\nu_i-1)\sum_l\sum_{\pi\in S_{N-2}}
\pi\ket{\dots \underset{\underset{l}{\downarrow}}{s}^is^i\dots}\\
-2\sum_l\sum_{\substack{1\leq j,k\leq n\\j,k\neq i}}\sum_{\pi\in S_{N-2}}
\nu_j(\nu_k-\de_{jk})\pi\ket{\dots \underset{\underset{l}{\downarrow}}{s}^js^k\dots}\,,
\end{multline}
where the permutations $\pi$ act only on the $N-2$ spin components specified by the ellipses.
On the other hand, we have
\begin{multline}\label{Las}
N\cdot N!\La\ket{\bs}=\sum_l\sum_{\pi\in S_{N-2}}\nu_i(\nu_i-1)
\pi\ket{\dots \underset{\underset{l}{\downarrow}}{s}^is^i\dots}\\
{}+2\sum_l\sum_{\substack{1\leq j\leq n\\j\neq i}}\sum_{\pi\in S_{N-1}}
\nu_j\pi\ket{\dots \underset{\underset{l}{\downarrow}}{s}^j\dots}\\
-\sum_l\sum_{\substack{1\leq j,k\leq n\\j,k\neq i}}\sum_{\pi\in S_{N-2}}
\nu_j(\nu_k-\de_{jk})\pi\ket{\dots \underset{\underset{l}{\downarrow}}{s}^js^k\dots}\,.
\end{multline}
Comparing Eqs.~\eqref{Achia} and \eqref{Las} we obtain
\begin{align}
A\ket{\chi_i}&=2\bigg(N\La\ket\bs-\frac2{N!}\,\sum_{\substack{1\leq j\leq n\\j\neq i}}\nu_j\sum_l
\sum_{\pi\in S_{N-1}}\pi\ket{\dots \underset{\underset{l}{\downarrow}}{s}^j\dots}\bigg)\notag\\
&=2\bigg(N-2\sum_{\substack{1\leq j\leq n\\j\neq i}}\nu_j\bigg)\La\ket\bs
=2(2\nu_i-N)\La\ket\bs\,.\label{chiainSip}
\end{align}
This shows that any state of the form~\eqref{chia} belongs to $\Si'/{\sim}$.
Suppose next that $\nu_i=1$, so that $\ket{\chi_i}$ is given by Eq.~\eqref{chib}.
Since
\begin{equation}\label{chibinSip}
A\ket{\chi_i}=\frac2{N!}\,
\sum_l\sum_{\substack{1\leq j\leq n\\j\neq i}}\sum_{\pi\in S_{N-2}}
\nu_j\pi\big(\ket{\dots \underset{\underset{l}{\downarrow}}{s}^is^j\dots}
+\ket{\dots \underset{\underset{l}{\downarrow}}{s}^js^i\dots}\big)=4\La\ket\bs\,,
\end{equation}
it follows that in this case $\ket{\chi_i}$ is also in $\Si'/{\sim}$\,.
\end{proof}
\begin{remark}
Just as symmetric spin states, cf.~Eq.~\eqref{AsLas}, the states
$\ket{\chi_i}$ satisfy the relation
\begin{equation}\label{AchiLachi}
A\ket{\chi_i}=\frac2{N-1}\,\La\ket{\chi_i}\,.
\end{equation}
Indeed, if $\nu_i>1$, from Eqs.~\eqref{chia} and~\eqref{chiainSip} we have
\begin{align*}
\La\ket{\chi_i}&=\bigg[\nu_i(\nu_i-1)
+\sum_{\substack{1\leq j\leq n\\j\neq i}}\nu_j
-\sum_{\substack{1\leq j,k\leq n\\j,k\neq i}}\nu_j\nu_k\bigg]\La\ket{\bs}\\
&=\big[\nu_i(\nu_i-1)+N-\nu_i-(N-\nu_i)^2\big]\La\ket{\bs}\\
&=(N-1)(2\nu_i-N)\La\ket{\bs}=\frac{N-1}2\,A\ket{\chi_i}\,.
\end{align*}
On the other hand, if $\nu_i=1$ Eqs.~\eqref{chib} and~\eqref{chibinSip} imply that
\[
\La\ket{\chi_i}=2\bigg(\sum_{\substack{1\leq j\leq n\\j\neq i}}\nu_j\bigg)\La\ket{\bs}
=2(N-1)\La\ket{\bs}=\frac{N-1}2\,A\ket{\chi_i}\,.
\]
\end{remark}
\begin{example}\label{ex1}
We shall now present all the states of the form~\eqref{chis} for spin $1/2$. In this case,
up to a permutation the basic state $\ket\bs$ is given by
\begin{equation}\label{nu}
\ket\bs=
\ket{\overbrace{\vphantom{|}{+}\dots+}^{\nu}\overbrace{\vphantom{|}-\dots-}^{N-\nu}\:}\,.
\end{equation}
If $\nu$ is either $0$ or $N$, then $n=1$ and thus $\ket{\chi_1}$ is of the type~\eqref{chia}
and proportional to $\ket\bs$. If $\nu=1$, then $n=2$ and we can take (dropping inessential factors)
\[
\ket{\chi_1}=\frac12\,\big(\ket{{+-}\cdots}+\ket{{-+}\cdots}\big)\sim\ket{{+-}\cdots}\,,\qquad
\ket{\chi_2}=\ket{{--}\cdots}\,.
\]
Although the states $\ket{\chi_1}$ and $\ket{\chi_2}$ are linearly independent, the combination
$2\ket{\chi_1}+(N-2)\ket{\chi_2}$ is equivalent to a symmetric state. In the case
$\nu=N-1$ the states $\ket{\chi_i}$ are obtained from the previous ones by flipping the spins.
Finally, if $2\leq \nu\leq N-2$ then $n=2$ and the states $\ket{\chi_i}$ are now given by
\begin{equation}\label{chi1}
\ket{\chi_1}=-\ket{\chi_2}=\nu(\nu-1)\ket{{++}\cdots}
-(N-\nu)(N-\nu-1)\ket{{--}\cdots}\,.
\end{equation}
\end{example}
According to the previous example, for spin $1/2$ there are exactly $n-1$ independent
states of the form~\eqref{chis} associated to each basic state $\ket\bs$,
up to symmetric states. We shall see next that this fact actually holds for arbitrary
spin:
\begin{proposition}
Let $\ket\bs$ be a basic spin state. If $n$ is the number of distinct
components of $\bs$, there are exactly $n-1$ independent states of the
form~\eqref{chis} modulo symmetric states.
\end{proposition}
\begin{proof}
Let $p$ be the number of distinct components $s^i$ of $\bs$ such
that $\nu_i>1$. A straightforward computation shows that the
combination
\begin{equation}\label{sumchis}
\sum_{i=1}^n\ket{\chi_i}\sim(2-p)\sum_{i,j=1}^n\nu_i(\nu_j-\de_{ij})\ket{s^is^j\cdots}
\sim(2-p)N(N-1)\La\ket\bs
\end{equation}
is equivalent to a symmetric state. Suppose first that $p\neq2$. It is immediate
to check that in this case the set $\{\ket{\chi_i}\mid
i=1,\dots,n\}$ is linearly independent. If a linear combination
$\sum_{i=1}^nc_i\ket{\chi_i}$ is equivalent to a symmetric state
$\ket\hs$, then $\ket\hs$ must be proportional to $\La\ket\bs$, so
that we can write
\[
\sum_{i=1}^nc_i\ket{\chi_i}\sim\la(2-p)N(N-1)\La\ket\bs\,.
\]
Hence $\sum_{i=1}^n(c_i-\la)\ket{\chi_i}\sim0$, and the linear
independence of the states $\ket{\chi_i}$ implies that $c_i=\la$
for all $i$. On the other hand, if $p=2$ the set
$\{\ket{\chi_i}\mid i=1,\dots,n\}$ is linearly dependent on
account of Eq.~\eqref{sumchis}, but removing one of the two states
with $\nu_i>1$ clearly yields a linearly independent set. It is
also obvious from the coefficients of the states
$\ket{s^is^i\dots}$ that no linear combination
$\sum_{i=1}^nc_i\ket{\chi_i}$ can be equivalent to a nonzero
symmetric state.
\end{proof}
The next natural question to be addressed is whether the states of the
form~\eqref{chis} span the space $\Si'/{\sim}$ up to symmetric states:
\begin{proposition}\label{prop.span}
The space $\big(\Si'/\La(\Si)\big)/{\sim}$ is spanned
by states of the form~\eqref{chis}.
\end{proposition}
\begin{proof}
For conciseness, we present the proof of this result only for the case $M=1/2$.
Let $\Si_\nu$ denote the subspace of $\Si$ spanned by basic spin
states with $\nu$ ``${+}$'' spins, and set
$\Si'_\nu=\Si'\cap\Si_\nu$. Since the operators $A$ and $\La$
involved in the definition~\eqref{Sip} of $\Si'$ clearly preserve
$\Si_\nu$, it suffices to show that the states $\ket{\chi_i(\bs)}$
with $\bs$ given by~\eqref{nu} span the space $\Si'_\nu/{\sim}$ up
to symmetric states. Note first that the statement is trivial for
$\nu=0,1,N-1,N$, since in this case the states of the
form~\eqref{chis} obviously generate the whole space
$\Si_\nu/{\sim}$\,. Suppose, therefore, that $2\leq\nu\leq N-2$, so
that
\[
\Si_\nu/{\sim}=\big\langle\,\ket{{++}\cdots},\,\ket{{+-}\cdots},\,\ket{{--}\cdots}\,\big\rangle\,.
\]
Since the state~\eqref{chi1} and the symmetric state (up to
equivalence)
\[
\nu(\nu-1)\ket{{++}\cdots}+2\nu(N-\nu)\ket{{+-}\cdots}+(N-\nu)(N-\nu-1)\ket{{--}\cdots}
\]
both belong to $\Si'_\nu/{\sim}$, we need only show that (for
instance) $\ket{{+-}\cdots}$ is not in $\Si'_\nu/{\sim}$, i.e.,
that $A\ket{{+-}\cdots}$ is not symmetric. But this is certainly the
case, since a state of the form
\[
\ket{\overbrace{\vphantom{|}{+-}\cdots{+-}}^{2k}\ms
\overbrace{\vphantom{|}{-}\cdots-}^{N-\nu-k}\ms
\overbrace{\vphantom{|}{+}\cdots+}^{\nu-k}\:}\,,\qquad
k=1,2,\dots,\min(\nu,N-\nu)\,,
\]
appears in $A\ket{{+-}\cdots}$ with coefficient
$2k(\nu-1)!(N-\nu-1)!$ depending on $k$.
\end{proof}

\section{The algebraic eigenfunctions}\label{sec.eig}

In the previous section we have provided a detailed description of
the spaces $\cH^n_\ep\subset\La\big(\CC[\bz]\otimes\Si\big)$
invariant under the corresponding gauge Hamiltonian $\BH_\ep$. In
this section we shall explicitly compute all the eigenfunctions of
the restrictions of the operators $\BH_\ep$ to their invariant
spaces $\cH^{\ms n}_\ep$. This yields several infinite\footnote{For
the model~\eqref{V2} we shall see below that the number of
eigenfunctions with a given total momentum is finite.} families of
eigenfunctions for each of the models~\eqref{Vs}, which is the main
result of this paper. We shall use the term \emph{algebraic} to
refer to these eigenfunctions and their corresponding energies. It
is important to observe that the eigenfunctions of the gauge
Hamiltonian $\BH_\ep$ that can be constructed in this way are
necessarily invariant under the whole symmetric group, in spite of
the fact that $\BH_\ep$ is symmetric only under cyclic permutations.
In fact, the explicit solutions of all known CS models with
near-neighbors interactions (both in the scalar and spin cases) can
be factorized as the product of a simple gauge factor analogous to
$\mu$ times a completely symmetric
function~\cite{JK99,AJK01,DG01,EGKP05}. This, however, does not rule
out the existence of other eigenfunctions of the gauge Hamiltonian
$\BH_\ep$ invariant only under the subgroup of cyclic permutations,
which is indeed an interesting open problem.\bigskip

\ni{\bf Case~a}\medskip

We shall begin with the model~\eqref{V0}, which is probably
the most interesting one due to the rich structure of its associated invariant flag.
In order to find the algebraic energies of the model, note first that
one can clearly construct a basis $\cB_0^n$ of $\cH_0^n$ whose elements
are homogeneous polynomials in $\bz$ with coefficients in $\Si$. If
$F\in\cB_0^n$ has degree $k$, then $J^0F=kF$ and $T_0F$ has degree at most $k-2$.
If $\cB_0^n$ is ordered according to the degree, the operator $\BH_0$
is represented in this basis by a triangular matrix with diagonal elements
$E_0+kc_0$, where $k=0,\dots,n$ is the degree. Thus the algebraic energies
are the numbers
\[
E_k=E_0+2k\om,\qquad k=0,1,\dots.
\]

We shall next show that the algebraic eigenfunctions of $\BH_0$ can be expressed in
closed form in terms of generalized Laguerre and Jacobi polynomials. The
computation is basically a two-step procedure. In the first place, one encodes
the eigenvalue problem in the invariant space $\cH_0^n$ as a system of linear partial differential equations.
The second step then consists in finding the polynomial solutions of this system.

Regarding the first step, we shall need the following preliminary lemma:
\begin{lemma}
The operator $\BH_0$ preserves the following subspaces of $\cH_0^n$:
\begin{align}
& \BcH^n_{0,\ket s}=\langle f\Phi^{(0)},g\Phi^{(1)},h\Phi^{(2)}\rangle\,,
& &\ket s\in\Si\,,\label{BcH}\\
& \tcH^n_{0,\ket s}=\BcH^n_{0,\ket s}+\langle\Th\tPhi^{(2)}\rangle\,,
& &\ket s\in\Si'\,,\quad S_{12}\ket s=\ket s\,,\label{tcH}\\
& \hcH^n_{0,\ket s}=\BcH^n_{0,\ket s}+\langle w\hPhi^{(3)}\rangle\,,
& &\ket s\in\Si\,,\quad S_{12}\ket s=-\ket s\,,\label{hcH}
\end{align}
where $f$, $g$, $h$, $\Th$, $w$ are as in the definition of $\cT_0^n$ in Theorem~\ref{thm.1},
and $\Phi^{(k)}$, $\tPhi^{(2)}$, $\hPhi^{(3)}$ are respectively given
by~\eqref{Phik}, \eqref{tPhi2} and \eqref{hPhi3}.
\end{lemma}
\begin{proof}
The identity $T_0\big(f\Phi^{(0)}\big)=(T_0f)\Phi^{(0)}$ and
Eqs.~\eqref{Phi01}, \eqref{Phi02} and~\eqref{J0Phis} clearly imply
that the subspace $\BcH^n_{0,\ket s}$ is invariant under $\BH_0$.
Consider next the action of $\BH_0$ on a function of the form
$\Th\tPhi^{(2)}$.
Since $\ket s$ is symmetric under $S_{12}$, we can replace
$\La\big[(z_1+z_2)\ket s\big]$ by $2\Phi^{(1)}$ in
Eq.~\eqref{tPhi02}. Secondly, any state $\ket s\in\Si'$ satisfies
the identity
\begin{equation}\label{AsLas2}
\sum_i\ket{s^+_{i,i+1}}=\frac2{N-1}\,\La\ket s\,.
\end{equation}
Indeed, we already know that the previous identity holds for
symmetric states (cf.~Eq.~\eqref{AsLas}) and for states of the form~\eqref{chis}
(cf.~Eqs.~\eqref{A} and~\eqref{AchiLachi}). On the other hand, by Proposition~\ref{prop.span}
every state in $\Si'$ is a linear combination of a symmetric state, states of the
form~\eqref{chis}, and a state $\ket s$ such that \mbox{$\La\big(z_1z_2\ket s\big)=0$}.
But for the latter ``null'' state $\ket{s_{ij}}=0$ for all $i<j$, and hence $A\ket s=\La\ket s=0$.
Therefore, Eq.~\eqref{tPhi02} can be written as
\begin{equation}\label{tPhi02mod}
T_0\big(\Th\tPhi^{(2)}\big)
=(T_0\Th+8\Th_2)\tPhi^{(2)}+4\Th_1\Phi^{(1)}-\frac{4a}{N-1}\,\Th\Phi^{(0)}\,.
\end{equation}
{}From the previous equation and Eq.~\eqref{J0Phis} it follows
that $\BH_0\big(\Th\tPhi^{(2)}\big)\in\tcH^n_{0,\ket s}$.
Finally, if $S_{12}\ket s=-\ket s$, Eq.~\eqref{hPhi03} reduces to
\begin{equation}
T_0\big(w\hPhi^{(3)}\big)=(T_0w+12w_2)\hPhi^{(3)}+4w_1\Phi^{(2)}-4(2a+1)w\Phi^{(1)}\,,
\label{hPhi03mod}
\end{equation}
which, together with Eq.~\eqref{J0Phis}, implies that
$\BH_0\big(w\hPhi^{(3)}\big)\in\hcH^n_{0,\ket s}$.
\end{proof}
\begin{remark}
The requirement that $\ket s$ be symmetric (respectively antisymmetric) under $S_{12}$ in the
definition of the space $\tcH^n_{0,\ket s}$ (respectively $\hcH^n_{0,\ket s}$)
is no real restriction, since the antisymmetric (respectively symmetric)
part of $\ket s$ does not contribute to the state $\tPhi^{(2)}$ (respectively $\hPhi^{(3)}$).
\end{remark}

By the previous lemma, we can consider without loss of generality
eigenfunctions of $\BH_0$ of the form
\begin{equation}\label{Phi}
\Phi=f\Phi^{(0)}+g\Phi^{(1)}+h\Phi^{(2)}+\Th\tPhi^{(2)}+w\hPhi^{(3)}\,,\qquad\deg\Phi=k\,,
\end{equation}
where the spin functions $\Phi^{(k)}$, $\tPhi^{(2)}$ and $\hPhi^{(3)}$ are all
built from the same spin state $\ket s$. Note that we can assume
that $\Th\ms w=0$, and that the spin state $\ket s$ is symmetric under $S_{12}$ and
belongs to $\Si'$ if $\Th\neq 0$, whereas it is antisymmetric under $S_{12}$ if $w\neq 0$.

Using Eqs.~\eqref{Phi01}, \eqref{Phi02}, \eqref{J0Phis}, \eqref{tPhi02mod}
and~\eqref{hPhi03mod}, it is straightforward to show that the
eigenvalue equation $\BH_0\Phi=(E_0+2k\om )\Phi$ is equivalent to
the system
\begin{subequations}
\begin{align}
& \big[{-T_0}+2\om(J^0+3-k)\big]w-12w_2=0\,,\label{systemw}\\
& \big[{-T_0}+2\om(J^0+2-k)\big]\Th-8\Th_2=0\,,\label{systemTh}\\
& \big[{-T_0}+2\om(J^0+2-k)\big]h-8h_2=6g_3+4w_1\,,\label{systemh}\\
& \big[{-T_0}+2\om(J^0+1-k)\big]g-4g_2=4(h_1+\Th_1)-4(2a+1)w\,,\label{systemg}\\
& \big[{-T_0}+2\om(J^0-k)\big]f=2\Big(g_1+(2a+1)h-\frac{2a}{N-1}\,\Th\Big)\,.\label{systemf}
\end{align}
\end{subequations}
Since $f$ and $g$ are linear in $\si_3$, we can write
\begin{equation}\label{fg}
f=p+\si_3 q\,,\qquad g=u+\si_3 v\,,
\end{equation}
where $p$, $q$, $u$ and $v$ are polynomials in $\si_1$ and $\si_2$.
Taking into account that the action of $T_0$ on scalar symmetric
functions is given by the RHS of Eq.~\eqref{Tep-res} with $\ep=0$, and using
Eqs.~\eqref{X0f}, \eqref{T20f} and~\eqref{J0f0}, we finally obtain
the following linear system of PDEs:
\begin{subequations}\label{system2}
\begin{align}
& \big[L_0-2\om(k-3)\big]w-12w_2=0\,,\label{system2w}\\
& \big[L_0-2\om(k-2)\big]\Th-8\Th_2=0\,,\label{system2th}\\
& \big[L_0-2\om(k-2)\big]h-8h_2=6v+4w_1\,,\label{system2h}\\
& \big[L_0-2\om(k-1)\big]u-4u_2=4h_1+4\Th_1+6\si_2v_1\notag\\
& \hphantom{\big[L_0-2\om(k-1)\big]u-4u_2=4h_1}+6(2a+1)\si_1 v-4(2a+1)w\,,\label{system2u}\\
& \big[L_0-2\om(k-4)\big]v-16v_2=0\,,\label{system2v}\\
& \big(L_0-2\om k\big)p=2u_1+2(2a+1)h-\frac{4a}{N-1}\,\Th+6\si_2q_1+6(2a+1)\si_1q\,,\label{system2p}\\
& \big[L_0-2\om(k-3)\big]q-12q_2=2v_1\,,\label{system2q}
\end{align}
\end{subequations}
where
\begin{multline}\label{L0}
L_0=-\big(N\pa_{\si_1}^2+4\si_1\pa_{\si_1}\pa_{\si_2}+4\si_2\pa_{\si_2}^2+2(2a+1)N\pa_{\si_2}\big)
\\+2\om(\si_1\pa_{\si_1}+2\si_2\pa_{\si_2})\,.
\end{multline}
As a consequence of the general discussion of the previous Section,
the latter system is guaranteed to possess polynomials solutions.
In fact, these polynomial solutions can be expressed in closed form in terms
of generalized Laguerre polynomials $L^\la_\nu$ and Jacobi polynomials $P^{(\ga,\de)}_\nu$.
\begin{theorem}\label{thm.H0}
Let
\[
\al=N(a+\frac12)-\frac32\,,\qquad
\be\equiv\be(m)=1-m-N\Big(a+\frac12\Big)\,,\qquad
t=\frac{2r^2}{N\Bx^2}-1\,,
\]
where $\Bx=\frac 1N\sum_ix_i$ is the center of mass coordinate. The Hamiltonian $H_0$ possesses
the following families of spin eigenfunctions with eigenvalue \mbox{$E_{lm}=E_0+2\om(2l+m)$},
with $l\geq 0$ and $m$ as indicated in each case:
\begin{align*}
\Psi^{(0)}_{lm}&=\mu\ms\Bx^mL^{-\be}_l(\omega r^2)
P^{(\al,\be)}_{[\frac m2]}(t)\,\Phi^{(0)}\,,\qquad m\geq0\,,\\
\Psi^{(1)}_{lm}&=\mu\ms\Bx^{m-1}L^{-\be}_l(\om
r^2)P^{(\al+1,\be)}_{[\frac{m-1}2]}(t)\big(\Phi^{(1)}-\Bx \,\Phi^{(0)}\big)\,,\qquad m\geq1\,,\displaybreak[0]\\
\Psi^{(2)}_{lm}&=\mu\ms\Bx^{m-2}L^{-\be}_l(\om
r^2)\bigg[P^{(\al+2,\be)}_{[\frac m2]-1}(t)\big(\Phi^{(2)}-2\Bx\,\Phi^{(1)}\big)\\
&\qquad\qquad\qquad+\Bx^2\bigg(P^{(\al+2,\be)}_{[\frac
m2]-1}(t)-\frac{2(\al+1)}{2[\tfrac{m-1}2]+1}\,P^{(\al+1,\be)}_{[\frac
m2]-1}(t)\bigg)\Phi^{(0)}\bigg]\,,\quad m\geq2\,,\displaybreak[0]\\
\widetilde\Psi^{(2)}_{lm}&=\mu\ms\Bx^{m-2}L^{-\be}_l(\om
r^2)\bigg[P^{(\al+2,\be)}_{[\frac m2]-1}(t)\big(\widetilde\Phi^{(2)}-2\Bx\,\Phi^{(1)}\big)\\
&\qquad+\Bx^2\bigg(P^{(\al+2,\be)}_{[\frac
m2]-1}(t)+\frac{2(\al+1)}{\big(2[\tfrac{m-1}2]+1\big)(N-1)}\,P^{(\al+1,\be)}_{[\frac
m2]-1}(t)\bigg)\Phi^{(0)}\bigg]\,,\quad m\geq2\,,\displaybreak[0]\\
\Psi^{(3)}_{lm}&=\mu\ms\Bx^{m-3}L^{-\be}_l(\omega r^2)\bigg[
\frac2{3N}\,P^{(\al+3,\be)}_{[\frac{m-3}2]}(t)\sum_i
x_i^3+\Bx^3\vp_m(t)\bigg]\,\Phi^{(0)}\,, \qquad m\geq3\,,\displaybreak[0]\\
\hPsi^{(3)}_{lm}&=\mu\ms\Bx^{m-3}L^{-\be}_l(\omega r^2)\bigg[
P^{(\al+3,\be)}_{[\frac{m-3}2]}(t)\big(\hPhi^{(3)}-2\Bx\,\Phi^{(2)}\big)\\
&\qquad\qquad\qquad+2\Bx^2\bigg(P^{(\al+3,\be)}_{[\frac
{m-3}2]}(t)+\frac{2(\al+2)}{2[\frac m2]-1}\,P^{(\al+2,\be)}_{[\frac
{m-3}2]}(t)\bigg)\Phi^{(1)}\\
&\qquad\qquad\qquad-2\Bx^3\bigg(\frac13\,P^{(\al+3,\be)}_{[\frac{m-3}2]}(t)
+\frac1{2[\frac m2]-1}\,P^{(\al+2,\be)}_{[\frac{m-3}2]}(t)\\
&\qquad\qquad\qquad\qquad\qquad
+\frac{2\al+3}{m(m-2)}\,\vep(m)P^{(\al+1,\be)}_{\frac{m-3}2}(t)\bigg)\Phi^{(0)}\bigg]
\,,\quad m\geq3\,,\displaybreak[0]\\
\Psi^{(4)}_{lm}&=\mu\ms \Bx^{m-4}L^{-\be}_l(\om
r^2)\bigg[\frac3{2([\frac{m-3}2]+\frac12)}\,\Bx^2P^{(\al+3,\be)}_{[\frac
m2]-2}(t)\,\Phi^{(2)}\\
&\qquad\qquad\qquad+\Big(\frac32\,\Bx^3\phi_m(t)-\frac{1}N\,P^{(\al+4,\be)}_{[\frac
m2]-2}(t)\sum_i x_i^3\Big)\Phi^{(1)}\\
&\qquad\qquad\qquad+\Big(\frac1N\,\Bx P^{(\al+4,\be)}_{[\frac
m2]-2}(t)\sum_ix_i^3+\frac32\,\Bx^4\chi_m(t)\Big)\Phi^{(0)}\bigg]\,,\quad m\geq4\,.
\end{align*}
Here $[\ms\cdot\ms]$ denotes the integer part, $\vep(m)=\big(1-(-1)^m\big)/2$, and
\[
\Phi^{(k)}=\La(x_1^k\ket s),\qquad\tPhi^{(2)}=\La(x_1x_2\ket s),\qquad
\hPhi^{(3)}=\La(x_1x_2(x_1-x_2)\ket s),
\]
where the spin state $\ket s$ is symmetric under $S_{12}$ and belongs to $\Si'$
for the eigenfunction $\widetilde\Psi^{(2)}_{lm}$, and is antisymmetric under $S_{12}$
for the eigenfunction $\hPsi^{(3)}_{lm}$.
The functions $\vp_m$, $\phi_m$ and $\chi_m$ are polynomials given explicitly by
\begin{align*}
\varphi_m&=\frac{m+2\al+2}{m-1}\,P^{(\al+2,\be-2)}_{\frac m2}
-P^{(\al+3,\be-1)}_{\frac
m2-1}-\frac{4\al+7}{m-1}\,P^{(\al+2,\be-1)}_{\frac m2-1}
+\frac13\,P^{(\al+3,\be)}_{\frac m2-2}\,,\displaybreak[0]\\
\phi_m&=P^{(\al+4,\be-1)}_{\frac m2-1}-2P^{(\al+3,\be-1)}_{\frac
m2-1}-\frac{m+2\al+3}{(m-1)(m-3)}\,P^{(\al+2,\be-1)}_{\frac
m2-1}\\
&\hphantom{{}=P^{(\al+4,\be-1)}_{\frac m2-1}}-\frac13\,P^{(\al+4,\be)}_{\frac
m2-2}+\frac{m+2\al-1}{m-3}\,P^{(\al+3,\be)}_{\frac m2-2}\,,\displaybreak[0]\\
\chi_m&=\frac{3m+2\al}{(m-1)(m-3)}\,P^{(\al+2,\be-1)}_{\frac
m2-1}+\frac{2m-7}{m-3}\,P^{(\al+3,\be-1)}_{\frac
m2-1}-P^{(\al+4,\be-1)}_{\frac m2-1}\\
&\hphantom{{}=P^{(\al+4,\be-1)}_{\frac m2-1}}-\frac{m+2\al+2}{(m-1)(m-3)}\,P^{(\al+2,\be)}_{\frac
m2-2}-\frac{m+2\al}{m-3}\,P^{(\al+3,\be)}_{\frac
m2-2}+\frac13\,P^{(\al+4,\be)}_{\frac m2-2}\,,
\end{align*}
for even $m$, and
\begin{align*}
\varphi_m&=2P^{(\al+2,\be-1)}_{\frac{m-1}2}-P^{(\al+3,\be-1)}_{\frac{m-1}2}
+\frac13\,P^{(\al+3,\be)}_{\frac{m-3}2}
+\frac{m+2\al+2}{m(m-2)}\,P^{(\al+1,\be)}_{\frac{m-3}2}\\
&\hphantom{{}=2P^{(\al+2,\be-1)}_{\frac{m-1}2}}
-\frac{m+2\al+2}{m-2}\,P^{(\al+2,\be)}_{\frac{m-3}2}\,,\displaybreak[0]\\
\phi_m&=P^{(\al+4,\be-1)}_{\frac{m-3}2}-\frac{2m-5}{m-2}\,P^{(\al+3,\be)}_{\frac{m-3}2}-
\frac13\,P^{(\al+4,\be)}_{\frac{m-5}2}+\frac{m+2\al-1}{m-2}\,P^{(\al+3,\be)}_{\frac{m-5}2}\,,
\displaybreak[0]\\
\chi_m&=\frac{2m-3}{m(m-2)}\,P^{(\al+2,\be-1)}_{\frac{m-3}2}+\frac{2(m-3)}{m-2}
\,P^{(\al+3,\be-1)}_{\frac{m-3}2}-P^{(\al+4,\be-1)}_{\frac{m-3}2}\\
&\hphantom{{}=2P^{(\al+2,\be-1)}_{\frac{m-1}2}}-\frac{m+2\al+1}{m(m-2)}\,P^{
(\al+2,\be)}_{\frac{m-5}2}-\frac{m+2\al}{m-2}\,P^{(\al+3,\be)}_{\frac{m-5}2}
+\frac13\,P^{(\al+4,\be)}_{\frac{m-3}2}\,,
\end{align*}
for odd $m$. For every $n=0,1,\dots$, the above eigenfunctions with $2l+m\leq n$ span
the whole $H_0$-invariant space $\mu\ms\cH_0^n$.
\end{theorem}
\begin{proof}
Recall, to begin with, that the algebraic eigenfunctions of $H_0$ are of the form
$\Psi=\mu\Phi$, with $\mu$ given in Table~\ref{table:params} and $\Phi$
an eigenfunction of $\BH_0$ of the form~\eqref{Phi}-\eqref{fg}. In order to determine $\Phi$,
we must find the most general polynomial solution of the
linear system~\eqref{system2}. {}From the structure of this system it follows
that there are seven types of independent solutions, characterized by the vanishing
of certain subsets of the unknown functions $p,q,u,v,h,\Th,w$. These types are listed in
Table~\ref{table:sols0}, where in the last column we have indicated
the eigenfunction of $H_0$ obtained from each case.
\begin{table}[h]
\begin{center}
\caption{The seven types of solutions of the system~\eqref{system2} and
their corresponding eigenfunctions.}\label{table:sols0}
\begin{tabular}{lc}\hline
\vrule height 15pt depth 9pt width0pt {\hfill Conditions\hphantom{\qquad}\hfill}
& Eigenfunction\\ \hline
\BStrut $q=u=v=h=\Th=w=0,\quad p\neq0$\hphantom{\qquad} & $\Psi^{(0)}_{lm}$\\ \hline
\BStrut $u=v=h=\Th=w=0,\quad q\neq0$\hphantom{\qquad} & $\Psi^{(3)}_{lm}$\\ \hline
\BStrut $q=v=h=\Th=w=0,\quad u\neq0$\hphantom{\qquad} & $\Psi^{(1)}_{lm}$\\ \hline
\BStrut $q=v=\Th=w=0,\quad h\neq0$\hphantom{\qquad} & $\Psi^{(2)}_{lm}$\\ \hline
\BStrut $q=v=h=w=0,\quad \Th\neq0$\hphantom{\qquad} & $\widetilde\Psi^{(2)}_{lm}$\\ \hline
\BStrut $q=v=\Th=0,\quad w\neq0$\hphantom{\qquad} & $\hPsi^{(3)}_{lm}$\\ \hline
\BStrut $\Th=w=0,\quad v\neq0$\hphantom{\qquad} & $\Psi^{(4)}_{lm}$\\ \hline
\end{tabular}
\end{center}
\end{table}
We shall present here in detail the solution of the system~\eqref{system2}
for the case $q=v=h=\Th=w=0$ and $u\neq0$, which yields the eigenfunctions of the form
$\Psi^{(1)}_{lm}$ (the procedure for the other cases is essentially the same). In this
case the system~\eqref{system2} reduces to
\begin{equation}
\big[L_0-2\om(k-1)\big]u-4u_2=0\,,\qquad
\big(L_0-2\om k\big)p=2u_1\,.\label{system3}
\end{equation}
Let us begin with the homogeneous equation for $u$. We shall look for polynomial solutions
of this equation of the form $u=Q(\si_1,\si_2)R(\si_2)$,
where $Q$ is a homogeneous polynomial of degree $m-1$ in $\bz$ and $R$ is a polynomial
of degree $l$ in~$\si_2$, so that $k=\deg\Phi=2l+m$ by Eq.~\eqref{Phi}.
{}From Eq.~\eqref{L0} and the homogeneity of $Q$ we have
\begin{align*}
L_0(QR)&=(L_0Q)R+Q(L_0R)-4\si_1Q_1R_2-8\si_2Q_2R_2\\
&=(L_0Q)R+Q(L_0-4(m-1)\pa_{\si_2})R\,.
\end{align*}
Hence the equation for $u$ can be written as
\[
R\big(\widehat L_0-4\pa_{\si_2}\big)Q=Q\big(-L_0+4m\pa_{\si_2}+4l\om\big)R\,,
\]
where $\widehat L_0=L_0|_{\om=0}$. Since $\big(\widehat L_0-4\pa_{\si_2}\big)Q$
is a homogeneous polynomial of degree $m-3$ in $\bz$, both sides of the latter
equation must vanish separately. We are thus led to the following decoupled
equations for $Q$ and $R$:
\begin{align}
&\big(\widehat L_0-4\pa_{\si_2}\big)Q=0\label{eqQ}\,,\\
&\big(-L_0+4m\pa_{\si_2}+4l\om\big)R=0\label{eqR}\,.
\end{align}
In terms of the variable $\rho=\om\si_2$, Eq.~\eqref{eqR} can be written as
\[
4\om\cL^{-\be}_l(R)=0\,,
\]
where
\begin{equation}\label{Lop}
\cL^\la_\nu=\rho\ms\pa_\rho^2+(\la+1-\rho)\pa_\rho+\nu
\end{equation}
is the generalized Laguerre operator. Hence $R$ is proportional to the
generalized Laguerre polynomial $L^{-\be}_l(\om\si_2)$. On the other
hand, we can write $Q=\si_1^{m-1}P(t)$ where $P$ is a polynomial
in the homogeneous variable $t=\frac{2N\si_2}{\si_1^2}-1$.
With this substitution, Eq.~\eqref{eqQ} becomes
\[
4N\si_1^{m-3}\cJ^{(\al+1,\be)}_{[\frac{m-1}2]}(P)=0\,,
\]
where the Jacobi operator $\cJ^{(\ga,\de)}_\nu$ is given by
\[
\cJ^{(\ga,\de)}_\nu=(1-t^2)\ms\pa_t^2+\big[\de-\ga-(\ga+\de+2)\ms t\big]\ms\pa_t+\nu(\nu+\ga+\de+1)\,.
\]
Thus $P(t)$ is proportional to the Jacobi polynomial
$P^{(\al+1,\be)}_{[\frac{m-1}2]}(t)$, so that we can take
\begin{equation}\label{u}
u=\si_1^{m-1}L^{-\be}_l(\om\si_2)P^{(\al+1,\be)}_{[\frac{m-1}2]}(t)\,.
\end{equation}

We must next find a particular solution of the inhomogeneous equation for $p$ in~\eqref{system3},
since the general solution of the corresponding homogeneous equation yields an eigenfunction
of the simpler type $\Psi^{(0)}_{lm}$. Since
\[
u_1=\si_1^{m-2}L^{-\be}_l(\om\si_2)\Big[
(m-1)P^{(\al+1,\be)}_{[\frac{m-1}2]}(t)-2(t+1)\dot P^{(\al+1,\be)}_{[\frac{m-1}2]}(t)\Big]
\]
(where the dot denotes derivative with respect to $t$), we make the ansatz
$p=\BQ(\si_1,\si_2)\BR(\si_2)$, where $\BQ$ is a homogeneous polynomial of degree $m$ in $\bz$
and $\BR$ is a polynomial of degree $l$ in $\si_2$. Substituting this ansatz into the
second equation in~\eqref{system3} and proceeding as before we immediately obtain
\[
\BR\big(\widehat L_0\BQ\big)+\BQ\big(L_0-4m\pa_{\si_2}-4l\om\big)\BR=2u_1\,.
\]
If we set $\BR=L^{-\be}_l(\om\si_2)$ the second term of the LHS vanishes, and
cancelling the common factor $L^{-\be}_l(\om\si_2)$ we are left with the
following equation for $\BQ$:
\[
\widehat L_0\BQ=2\si_1^{m-2}\Big[
(m-1)P^{(\al+1,\be)}_{[\frac{m-1}2]}(t)-2(t+1)\dot P^{(\al+1,\be)}_{[\frac{m-1}2]}(t)\Big]\,.
\]
The form of the RHS of this equation suggests the ansatz $\BQ=\si_1^m\BP(t)$,
with $\BP$ a polynomial in the variable $t$. The previous equation then yields
\begin{equation}\label{cJBP}
\cJ^{(\al,\be)}_{[\frac m2]}(\BP)=\frac1{2N}\,\Big[
(m-1)P^{(\al+1,\be)}_{[\frac{m-1}2]}(t)-2(t+1)\dot P^{(\al+1,\be)}_{[\frac{m-1}2]}(t)\Big]\,.
\end{equation}
{}From the definition of the Jacobi operator we easily obtain
\[
\cJ^{(\al,\be)}_{[\frac m2]}=\cJ^{(\al+1,\be)}_{[\frac{m-1}2]}+(1+t)\,\pa_t
-\frac12\,(m-1)\,,
\]
which implies that $\ds\BP=-\frac1N\,P^{(\al+1,\be)}_{[\frac{m-1}2]}$ is a particular solution
of Eq.~\eqref{cJBP}. Hence
\begin{equation}\label{p}
p=-\frac1N\,\si_1^mL^{-\be}_l(\om\si_2)P^{(\al+1,\be)}_{[\frac{m-1}2]}(t)
\end{equation}
is a particular solution of the inhomogeneous equation in~\eqref{system3}.
We have thus shown that $\Phi=p\,\Phi^{(0)}+u\,\Phi^{(1)}$, with $u$ and $p$ respectively given by
Eqs.~\eqref{u} and~\eqref{p}, is an eigenfunction of $\BH_0$ with eigenvalue
$E_0+2\om(2l+m)$. Multiplying~$\Phi$ by the gauge factor $\mu$
we obtain the eigenfunction $\Psi^{(1)}_{lm}$ of $H_0$ in the statement.

It remains to show that the states $\Psi^{(k)}_{lm}$ ($k=0,\dots,4$),
$\widetilde\Psi^{(2)}_{lm}$ and $\hPsi^{(3)}_{lm}$ with \mbox{$2l+m\leq n$} generate
the spaces~\eqref{BcH}--\eqref{hcH}. Consider first the
``monomials'' of the form $\mu\si_1^m\si_2^l\Phi^{(0)}$, which belong to $\mu\ms\BcH^n_{0,\ket s}$
if $2l+m\leq n$. We can order such monomials as follows: we say that
$\mu\si_1^{m'}\si_2^{l'}\Phi^{(0)}\prec\mu\si_1^m\si_2^l\Phi^{(0)}$
if $2l'+m'<2l+m$, or $2l'+m'=2l+m$ and $m'<m$. {}From the expansion~\cite[Eq.~8.962.1]{GR00}
\[
P^{(\ga,\de)}_\nu(t)=\frac1{\nu!}\,\sum_{k=0}^\nu\frac1{2^kk!}\,
(-\nu)_k(\ga+\de+\nu+1)_k(\ga+k+1)_{\nu-k}\,(1-t)^k\,,
\]
where $(x)_k$ is the Pochhammer symbol
\[
(x)_k=x(x+1)\cdots(x+k-1)\,,
\]
it follows that $P^{(\ga,\de)}_\nu(0)>0$ provided that $\ga+1>0$ and $\ga+\de+2\nu<0$.
In particular, $P^{(\al,\be)}_{[\frac m2]}(0)>0$ since
\[
\al+1=N\bigg(a+\frac12\bigg)-\frac12>N-\frac12>0\,,\quad
\al+\be+2\bigg[\frac m2\bigg]=2\bigg[\frac m2\bigg]-m-\frac12\leq-\frac12<0\,.
\]
Hence we can write
\[
\Psi^{(0)}_{lm}=\mu\Phi^{(0)}\big(c_{lm}\si_1^m\si_2^l+\text{l.o.t.}\big)\,,
\]
where $c_{lm}\neq 0$, so that
\[
\big\langle\Psi^{(0)}_{lm} \mid 2l+m\leq n\big\rangle
=\big\langle\mu\si_1^m\si_2^l\Phi^{(0)} \mid 2l+m\leq n\big\rangle\,.
\]
Likewise, a similar argument shows that for $m\geq 1$
\[
\big\langle \mu\ms\Bx^{m-1}L^{-\be}_l(\om
r^2)P^{(\al+1,\be)}_{[\frac{m-1}2]}(t)\,\Phi^{(1)}\mid 2l+m\leq n\big\rangle
=\big\langle\mu\si_1^{m-1}\si_2^l\Phi^{(1)} \mid 2l+m\leq n\big\rangle\,,
\]
and therefore
\[
\big\langle\Psi^{(0)}_{lm},\,\Psi^{(1)}_{lm} \mid 2l+m\leq n\big\rangle
=\big\langle\mu\si_1^m\si_2^l\Phi^{(0)},\,\mu\si_1^{m-1}\si_2^l\Phi^{(1)}
\mid 2l+m\leq n\big\rangle\,.
\]
Proceeding in the same way with the remaining spin eigenfunctions we can finally show that
\[
\big\langle\Psi^{(k)}_{lm}\mid k=0,\dots,4\,,\,2l+m\leq n\big\rangle
=\mu\ms\BcH^n_{0,\ket s}\,,
\]
and that
\begin{align*}
&\mu\ms\BcH^n_{0,\ket s}+\big\langle\widetilde\Psi^{(2)}_{lm}\mid 2l+m\leq n\big\rangle
=\mu\ms\tcH^n_{0,\ket s}\,,\qquad\ket s\in\Si',\quad S_{12}\ket s=\ket s,\\
&\mu\ms\BcH^n_{0,\ket s}+\big\langle\hPsi^{(3)}_{lm}\mid 2l+m\leq n\big\rangle
=\mu\ms\hcH^n_{0,\ket s}\,,\qquad S_{12}\ket s=-\ket s,
\end{align*}
as claimed.
\end{proof}
\begin{remark}
By Remark~1, the coefficients of $\Phi^{(0)}$ in the spin eigenfunctions
$\Psi^{(0)}_{lm}$ and $\Psi^{(3)}_{lm}$ yield the two families of eigenfunctions
of the scalar reduction $H^{\text{sc}}_0$ of the model~\eqref{V0}
presented without proof in our previous paper~\cite{EFGR05b}. Earlier work
on the scalar model $H^{\text{sc}}_0$ had established the existence of two families
of eigenfunctions of the form $\mu L_l^{-\be}(\om r^2)p_\nu(\bx)$,
with $p_\nu$ a homogeneous polynomial of degree $\nu\geq 3$, only for $\nu\leq 6$ and
$N\geq\nu$~\cite{AJK01}. More recently, Ezung et~al.~\cite{EGKP05} have rederived
a very small subset of these scalar eigenfunctions by mapping $H^{\text{sc}}_0$ to $N$
decoupled oscillators.
\end{remark}\goodbreak

\ni{\bf Case~b}\medskip

Since $c_0=4\om$ in this case, reasoning as
before we conclude that the algebraic energies are the numbers
\[
E_k=E_0+4k\om,\qquad k=0,1,\dots,
\]
where $k$ is the degree in $\bz$ of the
corresponding eigenfunctions of $\BH_1$. We shall see below that
these eigenfunctions can be written in terms of generalized Laguerre
polynomials. To this end, we begin by identifying certain subspaces
of $\cH_1^n$ invariant under $\BH_1$.

\begin{lemma}
For any given spin state $\ket s\in\Si$, the operator $\BH_1$ preserves
the subspace
\begin{equation}\label{BcH1n}
\cH^n_{1,\ket s}=
\big\langle f(\si_1,\si_2)\,\Phi^{(0)},\,g(\si_1)\,\Phi^{(1)}\mid{}f_{22}=0\big\rangle
\subset\cH_1^n\,,
\end{equation}
where $f$ and $g$ are polynomials of degrees at most $n$ and $n-1$ in $\bz$,
respectively, and $\Phi^{(k)}$ is given by~\eqref{Phik}.
\end{lemma}

\begin{proof}
The statement follows from the obvious identity
$T_1\big(f\Phi^{(0)}\big)=(T_1f)\Phi^{(0)}$ and
Eqs.~\eqref{X1f}, \eqref{Tep-res}, \eqref{T21f}, \eqref{Phi11}, \eqref{J0Phis} and~\eqref{JmPhis}.
\end{proof}

By the previous lemma we can assume that the eigenfunctions of $\BH_1$ in $\cH^n_{1,\ket s}$
are of the form
\begin{equation}\label{Phib}
\Phi=f\Phi^{(0)}+g\Phi^{(1)}\,,\qquad\deg\Phi=k\leq n\,.
\end{equation}
{}From Eqs.~\eqref{Phi11}, \eqref{J0Phis} and \eqref{JmPhis} it easily follows that the eigenvalue
equation $\BH_1\Phi=(E_0+4k\om)\Phi$ can be cast into the system
\begin{subequations}\label{systemb}
\begin{align}
& \Big[{-T_1}+\om(J^0+1-k)-\Big(b+\frac12\Big)J^-\Big]g-2g_1=0\,,\label{systembg}\\
& \Big[{-T_1}+\om(J^0-k)-\Big(b+\frac12\Big)J^-\Big]f=\Big(2a+b+\frac12\Big)g\,.\label{systembf}
\end{align}
\end{subequations}
Since $f$ is linear in $\si_2$ (cf.~Eq.~\eqref{BcH1n}), we can write
\begin{equation}\label{fpq}
f=p+\si_2 q,
\end{equation}
where $p$ and $q$ are polynomials in $\si_1$. Using
Eqs.~\eqref{X1f}, \eqref{Tep-res}, \eqref{T21f}, \eqref{J0f1} and~\eqref{Jmf1} we
can easily rewrite the system~\eqref{systemb} as follows:
\begin{subequations}\label{systemb2}
\begin{align}
& \big[L_1-\om(k-1)\big]g-2g_1=0\,,\label{systemb2g}\\
& \big[L_1-\om(k-2)\big]q-4q_1=0\,,\label{systemb2q}\\
& \big(L_1-\om k\big)p=\Big(2a+b+\frac12\Big)g+2\Big(4a+b+\frac32\Big)\si_1q\,,\label{systemb2p}
\end{align}
\end{subequations}
where
\begin{equation}\label{L1}
L_1=-\si_1\pa_{\si_1}^2+\Big[\om\si_1-\Big(2a+b+\frac12\Big)N\Big]\pa_{\si_1}\,.
\end{equation}
The last step is to construct the polynomials solutions of the system~\eqref{systemb2},
which can be expressed in terms of generalized Laguerre polynomials, according to the following
theorem.
\begin{theorem}\label{thm.H1}
The Hamiltonian $H_1$ possesses the following families of spin eigenfunctions
with eigenvalue $E_k=E_0+4k\om$:
\begin{align*}
\Psi^{(0)}_k&=\mu L^{\al-1}_k(\omega r^2)\,\Phi^{(0)}\,,\qquad k\geq0\,,\\[1mm]
\Psi^{(1)}_k&=\mu L^{\al+1}_{k-1}(\omega r^2)\big[N\Phi^{(1)}-r^2\Phi^{(0)}\big]
\,,\qquad k\geq1\,,\\[1mm]
\Psi^{(2)}_k&=\mu L^{\al+3}_{k-2}(\omega r^2)\Big[N(\al+1)\sum_i x_i^4-\be\ms r^4\Big]\,\Phi^{(0)}\,,
\qquad k\geq 2\,,
\end{align*}
where $\al=N(2a+b+\frac12)$, $\be=N(4a+b+\frac32)$, and $\Phi^{(j)}=\La\big(x_1^{2j}\ket s\big)$,
with $j=0,1$ and $\ket s\in\Si$. For each $\ket s\in\Si$ and $n=0,1,\dots$, the above eigenfunctions
with $k\leq n$ span the whole $H_1$-invariant space $\mu\cH^n_{1,\ket s}$.
\end{theorem}

\begin{proof}
As in the previous case, the algebraic eigenfunctions of $H_1$ are of the form $\Psi=\mu\Phi$,
where $\mu$ is given in Table~\ref{table:params} and $\Phi$ is an eigenfunction of $\BH_1$
of the form~\eqref{Phib}-\eqref{fpq}. The functions $p$, $q$ and $g$ are polynomials in $\si_1$
determined by the system~\eqref{systemb2}, which in terms of the variable $t=\om\si_1\equiv\om r^2$
can be written as
\begin{equation}\label{systemb3}
\cL^{\al+1}_{k-1}g=\cL^{\al+3}_{k-2}q=0\,,\qquad
\cL^{\al-1}_kp=-\frac\al{N\om}\,g-\frac{2\be}{N\om^2}\,tq\,,
\end{equation}
where $\cL^{\la}_\nu$ is the generalized Laguerre operator (cf. Eq.~\eqref{Lop}).
The general polynomial solutions of the first two equations in~\eqref{systemb3}
are respectively given by
\begin{equation}\label{gq}
g=c_1L^{\al+1}_{k-1}(t),\qquad q=c_2L^{\al+3}_{k-2}(t).
\end{equation}
On the other hand, from the elementary identity
\[
\cL^{\la}_\nu\Big(t^lL^{\la+2l}_{\nu-l}(t)\Big)=l(l+\la)t^{l-1}L^{\la+2l}_{\nu-l}(t)\,,
\]
it follows that the general polynomial solution of the third equation in~\eqref{systemb3}
is given by
\begin{equation}\label{bp}
p=c_0\ms L^{\al-1}_k(t)-\frac{c_1}{N\om}\,t\ms L^{\al+1}_{k-1}(t)-\frac{c_2\be}{N\om^2(\al+1)}\,t^2 L^{\al+3}_{k-2}(t)\,.
\end{equation}
Equations~\eqref{gq} and~\eqref{bp} immediately yield the formulas
of the eigenfunctions in this case. The last assertion in the
statement of the theorem follows from the fact that the functions
$p$, $q$ and $g$ in Eqs.~\eqref{gq} and~\eqref{bp} are the most
general polynomial solution of the system~\eqref{systemb3}.
\end{proof}

\begin{remark}
The spin eigenfunctions $\Psi^{(j)}_k$, $j=0,1,2$, listed in the previous theorem
are essentially the same as those presented in Ref.~\cite{EFGR05b} (note that in
the latter reference there is a typo in the formula for the scalar eigenfunction
$\psi^{(1)}_n$, namely the coefficient $\al$ multiplying $r^4$ should be replaced by
the parameter $\be$ defined in Theorem~\ref{thm.H1}).
\end{remark}

\begin{remark}
It should be noted that for $\om=0$ the potentials~\eqref{V0}
and~\eqref{V1} scale as $r^{-2}$ under dilations of the coordinates
(as is the case for the original Calogero model). The standard
argument used in the solution of the Calogero model shows that there
is a basis of eigenfunctions of these models of the form $\mu(\bx)
L^\la_\nu(\om r^2)F(\bx)$, where $F$ is a homogeneous spin-valued
function. The eigenfunctions presented in Theorems~\ref{thm.H0}
and~\ref{thm.H1} are indeed of this form.
\end{remark}

\ni{\bf Case~c}\medskip

The model~\eqref{V2} is of less interest than
the previous ones, since we shall see that in this case the number
of independent algebraic eigenfunctions is essentially finite. We
shall take, for definiteness, the plus sign in the change of
variable listed in Table~\ref{table:params} (it will be apparent
from the discussion that follows that the minus sign does not yield
additional solutions).

Let us first note that the potential for this model is
translationally invariant, so that the total momentum operator
$P=-\iu\sum_k\pa_{x_k}$ commutes with the Hamiltonian $H_2$. Hence
the eigenfunctions of $H_2$ can be assumed to have well-defined
total momentum. Equivalently, since
\begin{equation}\label{PJ0}
{\mu^{-1}\cdot P\cdot\mu\ms\Big|}_{x_k=-\frac\iu2\ms\log{z_k}}=2J^0,
\end{equation}
the eigenfunctions of $\BH_2$ can be assumed to be homogeneous in $\bz$.
Let $\Phi$ be a homogeneous eigenfunction of $\BH_2$ of degree $k$ and eigenvalue $E$,
so that $\Psi=\mu\Phi$ is an eigenfunction of $H_2$ with total momentum $2k$
(cf.~Eq.~\eqref{PJ0}) and energy $E$. By Eq.~\eqref{PJ0}, the function $\tau_N^\la\Psi$
clearly has total momentum $2(k+N\la)$. In fact, the following lemma implies that
$\tau_N^\la\Psi$ is also an eigenfunction of $H_2$ with a suitably boosted energy:
\begin{lemma}\label{lemma.5}
Let $\Phi$ be a homogeneous eigenfunction of $\BH_2$ of degree $k$ and eigenvalue~$E$. Then
$\tau_N^\la\Phi$ is an eigenfunction of $\BH_2$ with eigenvalue $E+8k\la+4N\la^2$.
\end{lemma}
\begin{proof}
{}From the identity
\[
\sum_i\frac1{z_i-z_{i+1}}\,\big(z_i^2\pa_i-z_{i+1}^2\pa_{i+1}\big)
=J^0+\frac12\sum_i\frac{z_i+z_{i+1}}{z_i-z_{i+1}}\,(D_i-D_{i+1})\,,
\]
where $D_i=z_i\pa_i$, we immediately obtain the following expression for the gauge
Hamiltonian $\BH_2$:
\begin{multline}
\frac14\,(\BH_2-E_0)=\sum_iD_i^2+a\sum_i\frac{z_i+z_{i+1}}{z_i-z_{i+1}}\,(D_i-D_{i+1})\\
-2a\sum_i\frac{z_iz_{i+1}}{(z_i-z_{i+1})^2}\,(1-K_{i,i+1})\,.
\end{multline}
Since $\tau_N^{-\la}D_i\tau_N^\la=D_i+\la$ for any real $\la$, it follows that
\[
\tau_N^{-\la}\BH_2\tau_N^\la=\BH_2+8\la J^0+4N\la^2\,.
\]
Taking into account that $J^0\Phi=k\Phi$, we conclude that
\[
\BH_2\big(\tau_N^\la\Phi\big)=\big(E+8k\la+4N\la^2\big)(\tau_N^\la\Phi)\,,
\]
as claimed.
\end{proof}
By the previous discussion, in what follows any two eigenfunctions of $\BH_2$
that differ by a power of $\tau_N$ shall be considered equivalent. {}From
Theorem~\ref{thm.1} and Corollary~\ref{cor.1} it easily follows that in this case
the number of independent algebraic eigenfunctions is finite, up to equivalence.
More precisely:
\begin{lemma}\label{lemma.6}
Up to equivalence, the algebraic eigenfunctions of $\BH_2$ can be assumed to belong to
a space of the form
\begin{equation}\label{BcH2kets}
\cH_{2,\ket s}=\langle\si_1,\,\tau_{N-1},\,\si_1\tau_{N-1},\,\tau_N\rangle\,\Phi^{(0)}
+\langle 1,\,\tau_{N-1}\rangle\,\Phi^{(1)}
+\langle 1,\,\si_1\rangle\,\tau_N\Phi^{(-1)}
\end{equation}
for some spin state $\ket s$, where $\Phi^{(k)}$ is given by~\eqref{Phik}.
\end{lemma}
\begin{proof}
Given a spin state $\ket s$, the obvious identity
\begin{equation}\label{T2fPhi0}
T_2\big(f\Phi^{(0)}\big)=(T_2f)\Phi^{(0)}
\end{equation}
and Eqs.~\eqref{BHep}, \eqref{Phi21}, \eqref{Phi2-1}, and~\eqref{J0Phis}
imply that the gauge Hamiltonian $\BH_2$ preserves the space
\begin{equation}\label{BcH2n}
\cH^n_{2,\ket s}=\big\langle f\Phi^{(0)},g\ms\Phi^{(1)},
\tau_Nq\ms\Phi^{(-1)}\big\rangle\,,
\end{equation}
where $f$, $g$ and $q$ are as in the definition of $\cT_2^n$ in
Theorem~\ref{thm.1}. Let $\Phi\in\cH^n_{2,\ket s}$ be an eigenfunction of $\BH_2$,
which as explained above can be taken as a homogeneous function of $\bz$.
{}From the conditions satisfied by the functions $f$, $g$ and $q$
in~\eqref{BcH2n} and the homogeneity of $\Phi$, it readily follows that
$\Phi\in\tau_N^l\cH_{2,\ket s}$ for some $l$, as claimed.
\end{proof}
\begin{theorem}\label{thm.H2}
The Hamiltonian $H_2$ possesses the following spin eigenfunctions
with zero momentum
\begin{gather*}
\Psi_0=\mu\,\Phi^{(0)},\qquad
\Psi_{1,2}=\mu\sum_i\bigg\{\begin{matrix}\cos\\\sin\end{matrix}\bigg\}
\big(2(x_i-\Bx)\big)\ket{s_i},\\
\Psi_3=\mu\bigg[\frac{2a}{2a+1}\,\Phi^{(0)}+\sum_{i\neq j}\cos\!\big(2(x_i-x_j)\big)\ket{s_j}\bigg],
\quad\Psi_4=\mu\sum_{i\neq j}\sin\!\big(2(x_i-x_j)\big)\ket{s_j},
\end{gather*}
where $\ket{s_i}$ is defined in~\eqref{sisij} and
$\Bx$ is the center of mass coordinate. Their energies are respectively given by
\[
E_0\,,\qquad E_{1,2}=E_0+4\Big(2a+1-\frac1N\Big)\,,\qquad E_{3,4}=E_0+8(2a+1)\,.
\]
Any algebraic eigenfunction with
well-defined total momentum is equivalent to a linear
combination of the above eigenfunctions.
\end{theorem}
\begin{proof} By Lemma~\ref{lemma.6}, in order to compute the algebraic eigenfunctions of $\BH_2$
it suffices to diagonalize $\BH_2$ in the spaces $\cH_{2,\ket s}$ given by~\eqref{BcH2kets}.
Equations~\eqref{Phi21}, \eqref{Phi2-1}, \eqref{J0Phis} and~\eqref{T2fPhi0}, and the fact
that $\BH_2$ preserves the degree of homogeneity, imply that the following subspaces of
$\cH_{2,\ket s}$ are invariant under $\BH_2$:
\begin{subequations}\label{invsubs2}
\begin{align}
& \langle\si_1\tau_{N-1}\,,\tau_N\rangle\,\Phi^{(0)}\,,\hspace*{-2cm}\label{invsubs20}\\
& \langle\si_1\ms\Phi^{(0)}\rangle\,, & &
\langle \si_1\ms\Phi^{(0)},\Phi^{(1)}\rangle\,, & &
\langle\si_1\tau_{N-1}\ms\Phi^{(0)},\tau_N\ms\Phi^{(0)},\tau_{N-1}\ms\Phi^{(1)}\rangle\,\,,\label{invsubs21}\\
& \langle\tau_{N-1}\ms\Phi^{(0)}\rangle\,, & &
\langle \tau_{N-1}\ms\Phi^{(0)},\tau_N\Phi^{(-1)}\rangle\,, & &
\langle\si_1\tau_{N-1}\ms\Phi^{(0)},\tau_N\ms\Phi^{(0)},\si_1\tau_N\ms\Phi^{(-1)}\rangle\,.\label{invsubs2-1}
\end{align}
\end{subequations}
{}From Eqs.~\eqref{invsubs2} it follows that the alternative
change of variables $z_k=\e^{-2\iu x_k}$ does not yield additional
eigenfunctions of $H_2$. Indeed, the latter change corresponds to
the mapping $z_k\mapsto 1/z_k$, which up to equivalence leaves the
subspace~\eqref{invsubs20} invariant and exchanges each subspace
in~\eqref{invsubs21} with the corresponding one
in~\eqref{invsubs2-1}. For this reason, we can safely ignore the
subspaces~\eqref{invsubs2-1} in the computation that follows,
provided that we add to the eigenfunctions of $\BH_2$ obtained from
the subspaces~\eqref{invsubs21} their images under the mapping
$z_k\mapsto 1/z_k$.

For the subspaces~\eqref{invsubs20}-\eqref{invsubs21},
using Eqs.~\eqref{BHep}, \eqref{Tep-res}, \eqref{T22f}, \eqref{Phi21}, \eqref{J0Phis},
and \eqref{J0f2} we easily obtain the following eigenfunctions of $\BH_2$:
\begin{subequations}\label{eigenf2}
\begin{align}
& \tau_N\ms\Phi^{(0)}, & & \quad E=E_0+4N,\label{eigenf2a}\\
& \Phi^{(1)}, & & \quad E=E_0+4(2a+1),\label{eigenf2b}\\
& \tau_{N-1}\ms\Phi^{(1)}-\frac{\tau_N}{2a+1}\,\Phi^{(0)}, & &\quad E=E_0+4(N+4a+2).\label{eigenf2c}
\end{align}
\end{subequations}
We have omitted the two additional eigenfunctions
\[
\si_1\ms\Phi^{(0)}\,,\qquad
\Big(\si_1\tau_{N-1}-\frac{N\tau_N}{2a+1}\Big)\ms\Phi^{(0)}
\]
from the above list, since they are respectively obtained from~\eqref{eigenf2b}
and~\eqref{eigenf2c} when the spin state $\ket s$ is symmetric. The eigenfunctions~\eqref{eigenf2}
are equivalent to the following ``zero momentum'' eigenfunctions:
\begin{subequations}\label{eigenf2zero}
\begin{align}
& \Phi^{(0)}, & & \quad E=E_0,\label{eigenf2zeroa}\\
& \tau_N^{-1/N}\Phi^{(1)}, & & \quad E=E_0+4\Big(2a+1-\frac1N\Big),\label{eigenf2zerob}\\
& \frac{\tau_{N-1}}{\tau_N}\,\Phi^{(1)}-\frac1{2a+1}\,\Phi^{(0)}, & &\quad
E=E_0+8(2a+1),\label{eigenf2zeroc}
\end{align}
\end{subequations}
where the energies have been computed from those in Eqs.~\eqref{eigenf2} using Lemma~\ref{lemma.5}.
The eigenfunctions of $H_2$ listed in the statement are readily obtained from these
spin functions together with the transforms of~\eqref{eigenf2zerob} and~\eqref{eigenf2zeroc}
under the mapping $z_k\mapsto 1/z_k$.
\end{proof}
\begin{remark}
If the spin state $\ket s$ is symmetric, then $\ket{s_i}=\Phi^{(0)}/N$ for all $i$, and
one easily obtains from Theorem~\ref{thm.H2} the following eigenfunctions of the scalar Hamiltonian
$H^{\mathrm{sc}}_2$:
\[
\psi_0=\mu,\quad
\psi_{1,2}=\mu\sum_i\bigg\{\begin{matrix}\cos\\\sin\end{matrix}\bigg\}
\big(2(x_i-\Bx)\big),\quad
\psi_3=\mu\bigg[\frac
{aN}{2a+1}+\sum_{i<j}\cos\!\big(2(x_i-x_j)\big)\bigg].
\]
These formulas agree with those in Refs.~\cite{EFGR05b} and~\cite{EGKP05}
(the expression of $\psi_3$ in the former reference contains an obvious erratum,
while this eigenfunction is missing altogether in the latter reference).
\end{remark}

\section{Summary and outlook}\label{sec.summ}

In this paper we have computed in closed form several infinite
families of eigenfunctions of the spin models with near-neighbors
interactions~\eqref{Vs} introduced in our previous
paper~\cite{EFGR05b}. Our method is based on the fact that each
spin Hamiltonian $H_\ep$ is related to a scalar operator $\BH_\ep$
involving difference operators which exchange pairs of neighboring
particles, cf.~Eqs.~\eqref{Tep}--\eqref{Hepstar}. We have
explicitly constructed a flag of finite-dimensional polynomial
subspaces $\cH^0_\ep\subset\cH^1_\ep\subset\cdots$ invariant
under $\BH_\ep$ (see Corollary~\ref{cor.1}). For all three
models~\eqref{Vs}, we have been able to fully diagonalize the
gauge Hamiltonian $\BH_\ep$ in its invariant spaces $\cH^n_\ep$.
Multiplying each eigenfunction of $\BH_\ep$ in $\cH^n_\ep$ by the
appropriate gauge factor $\mu$ and performing a suitable change of
variables (cf.~Table~\ref{table:params}) we obtain the families of
eigenfunctions of $H_\ep$ mentioned above.

The results obtained in this paper suggest several open problems
that we shall now briefly discuss. In the first place,
it would be natural to study the $BC_N$ counterparts
of the models~\eqref{Vs}, for which the interaction potential
also depends on the sums $x_i+x_{i+1}$. In fact, in the scalar case
this question has already been addressed in Ref.~\cite{AJK01}.
It would also be of interest to construct solvable models
with near-neighbors interactions of elliptic type, both in the
scalar and spin cases; see Refs.~\cite{FGGRZ01,FGGRZ01b} for
a list of models of this type with long-range interactions.
An important problem closely related with the subject of this paper
is the analysis of the spin chains obtained from the models~\eqref{Vs}
by applying the freezing trick. These chains are characterized by the fact that
the interactions are restricted to nearest neighbors (as in the Heisenberg chain),
but their strength depends on the distance between the sites
(as in chains of Haldane--Shastry type). For this reason, we believe
that the study of these new chains could prove of considerable interest.

Consider, for instance, the chain
associated with the model~\eqref{V0}, whose Hamiltonian is given by
\begin{equation}\label{ssH0}
\ssH_0=\sum_i(\xi_i-\xi_{i+1})^{-2}\ms S_{i,i+1},
\end{equation}
where $(\xi_1,\dots,\xi_N)$
is the unique equilibrium of the scalar potential
\[
U_0=\frac12\,r^2+\sum_i\frac1{(x_i-x_{i-1})(x_i-x_{i+1})}+
\sum_i\frac1{(x_i-x_{i+1})^2}
\]
in the domain $x_1<\cdots<x_N$. It can be shown that the chain sites $\xi_i$ are
symmetrically distributed around the origin; for instance, for $N=4$ we have
$\xi_4=-\xi_1=(\sqrt 3+1)/2$, $\xi_3=-\xi_2=(\sqrt 3-1)/2$. Note, in particular,
that the chain sites are not equally spaced, as is the case in most spin
chains of Haldane--Shastry type. In principle, it is not possible to apply the
method of Refs.~\cite{Po94,EFGR05,FG05} to evaluate the partition function of
the chain~\eqref{ssH0} in closed form, since the algebraic eigenfunctions of
the model~\eqref{V0} computed in Section~\ref{sec.eig} do not form a complete
set. On the other hand, the explicit nature of the algebraic eigenfunctions
presented in Theorem~\ref{thm.H0} makes it feasible to compute a number of
eigenvalues and eigenfunctions of the spin chain~\eqref{ssH0} by taking the
strong coupling limit $a\to\infty$. We emphasize that the results thus obtained would be
valid for an arbitrary number of spins, and thus could be helpful in
uncovering general properties of the spectrum. By combining this approach with
numerical computations for fixed values of $N$, we expect to achieve a
reasonable understanding of the properties of this novel type of chains.

\begin{acknowledgments}
  This work was partially supported by the DGI under grant no.~FIS2005-00752.
  A.E. acknowledges the financial support of the Spanish Ministry of Education
  through an FPU scholarship.
\end{acknowledgments}
%
%

\end{document}